\documentclass[11pt]{article}

\usepackage{fullpage,url}
\newenvironment{descit}[1]{\begin{quote} \textit{#1}}{\end{quote}}

\ifx\undefined\psfig\else \fi

\edef\psfigRestoreAt{\catcode`@=\number\catcode`@\relax}
\catcode`\@=11\relax
\newwrite\@unused
\def\typeout#1{{\let\protect\string\immediate\write\@unused{#1}}}
\def\figurepath{./}

\def\@nnil{\@nil}
\def\@empty{}
\def\@psdonoop#1\@@#2#3{}
\def\@psdo#1:=#2\do#3{\edef\@psdotmp{#2}\ifx\@psdotmp\@empty \else
    \expandafter\@psdoloop#2,\@nil,\@nil\@@#1{#3}\fi}
\def\@psdoloop#1,#2,#3\@@#4#5{\def#4{#1}\ifx #4\@nnil \else
       #5\def#4{#2}\ifx #4\@nnil \else#5\@ipsdoloop #3\@@#4{#5}\fi\fi}
\def\@ipsdoloop#1,#2\@@#3#4{\def#3{#1}\ifx #3\@nnil 
       \let\@nextwhile=\@psdonoop \else
      #4\relax\let\@nextwhile=\@ipsdoloop\fi\@nextwhile#2\@@#3{#4}}
\def\@tpsdo#1:=#2\do#3{\xdef\@psdotmp{#2}\ifx\@psdotmp\@empty \else
    \@tpsdoloop#2\@nil\@nil\@@#1{#3}\fi}
\def\@tpsdoloop#1#2\@@#3#4{\def#3{#1}\ifx #3\@nnil 
       \let\@nextwhile=\@psdonoop \else
      #4\relax\let\@nextwhile=\@tpsdoloop\fi\@nextwhile#2\@@#3{#4}}
\newread\ps@stream
\newif\ifnot@eof       % continue looking for the bounding box?
\newif\if@noisy        % report what you're making?
\newif\if@atend        % %%BoundingBox: has (at end) specification
\newif\if@psfile       % does this look like a PostScript file?
{\catcode`\%=12\global\gdef\epsf@start{%!}}
\def\epsf@PS{PS}
\def\epsf@getbb#1{%
\openin\ps@stream=#1
\ifeof\ps@stream\typeout{Error, File #1 not found}\else
   {\not@eoftrue \chardef\other=12
    \def\do##1{\catcode`##1=\other}\dospecials \catcode`\ =10
    \loop
       \if@psfile
	  \read\ps@stream to \epsf@fileline
       \else{
	  \obeyspaces
          \read\ps@stream to \epsf@tmp\global\let\epsf@fileline\epsf@tmp}
       \fi
       \ifeof\ps@stream\not@eoffalse\else
       \if@psfile\else
       \expandafter\epsf@test\epsf@fileline:. \\%
       \fi
          \expandafter\epsf@aux\epsf@fileline:. \\%
       \fi
   \ifnot@eof\repeat
   }\closein\ps@stream\fi}%
\long\def\epsf@test#1#2#3:#4\\{\def\epsf@testit{#1#2}
			\ifx\epsf@testit\epsf@start\else
\typeout{Warning! File does not start with `\epsf@start'.  It may not be a PostScript file.}
			\fi
			\@psfiletrue} % don't test after 1st line
{\catcode`\%=12\global\let\epsf@percent=%\global\def\epsf@bblit{%BoundingBox}}
\long\def\epsf@aux#1#2:#3\\{\ifx#1\epsf@percent
   \def\epsf@testit{#2}\ifx\epsf@testit\epsf@bblit
	\@atendfalse
        \epsf@atend #3 . \\%
	\if@atend	
	   \if@verbose{
		\typeout{psfig: found `(atend)'; continuing search}
	   }\fi
        \else
        \epsf@grab #3 . . . \\%
        \not@eoffalse
        \global\no@bbfalse
        \fi
   \fi\fi}%
\def\epsf@grab #1 #2 #3 #4 #5\\{%
   \global\def\epsf@llx{#1}\ifx\epsf@llx\empty
      \epsf@grab #2 #3 #4 #5 .\\\else
   \global\def\epsf@lly{#2}%
   \global\def\epsf@urx{#3}\global\def\epsf@ury{#4}\fi}%
\def\epsf@atendlit{(atend)} 
\def\epsf@atend #1 #2 #3\\{%
   \def\epsf@tmp{#1}\ifx\epsf@tmp\empty
      \epsf@atend #2 #3 .\\\else
   \ifx\epsf@tmp\epsf@atendlit\@atendtrue\fi\fi}

\chardef\letter = 11
\chardef\other = 12

\newif \ifdebug %%% turn me on to see TeX hard at work ...
\newif\ifc@mpute %%% don't need to compute some values
\c@mputetrue % but assume that we do

\let\then = \relax
\def\r@dian{pt }
\let\r@dians = \r@dian
\let\dimensionless@nit = \r@dian
\let\dimensionless@nits = \dimensionless@nit
\def\internal@nit{sp }
\let\internal@nits = \internal@nit
\newif\ifstillc@nverging
\def \Mess@ge #1{\ifdebug \then \message {#1} \fi}

{ %%% Things that need abnormal catcodes %%%
	\catcode `\@ = \letter
	\gdef \nodimen {\expandafter \n@dimen \the \dimen}
	\gdef \term #1 #2 #3%
	       {\edef \t@ {\the #1}%%% freeze parameter 1 (count, by value)
		\edef \t@@ {\expandafter \n@dimen \the #2\r@dian}%
		\t@rm {\t@} {\t@@} {#3}%
	       }
	\gdef \t@rm #1 #2 #3%
	       {{%
		\count 0 = 0
		\dimen 0 = 1 \dimensionless@nit
		\dimen 2 = #2\relax
		\Mess@ge {Calculating term #1 of \nodimen 2}%
		\loop
		\ifnum	\count 0 < #1
		\then	\advance \count 0 by 1
			\Mess@ge {Iteration \the \count 0 \space}%
			\Multiply \dimen 0 by {\dimen 2}%
			\Mess@ge {After multiplication, term = \nodimen 0}%
			\Divide \dimen 0 by {\count 0}%
			\Mess@ge {After division, term = \nodimen 0}%
		\repeat
		\Mess@ge {Final value for term #1 of 
				\nodimen 2 \space is \nodimen 0}%
		\xdef \Term {#3 = \nodimen 0 \r@dians}%
		\aftergroup \Term
	       }}
	\catcode `\p = \other
	\catcode `\t = \other
	\gdef \n@dimen #1pt{#1} %%% throw away the ``pt''
}

\def \Divide #1by #2{\divide #1 by #2} %%% just a synonym

\def \Multiply #1by #2%%% allows division of a dimen by a dimen
       {{%%% should really freeze parameter 2 (dimen, passed by value)
	\count 0 = #1\relax
	\count 2 = #2\relax
	\count 4 = 65536
	\Mess@ge {Before scaling, count 0 = \the \count 0 \space and
			count 2 = \the \count 2}%
	\ifnum	\count 0 > 32767 %%% do our best to avoid overflow
	\then	\divide \count 0 by 4
		\divide \count 4 by 4
	\else	\ifnum	\count 0 < -32767
		\then	\divide \count 0 by 4
			\divide \count 4 by 4
		\else
		\fi
	\fi
	\ifnum	\count 2 > 32767 %%% while retaining reasonable accuracy
	\then	\divide \count 2 by 4
		\divide \count 4 by 4
	\else	\ifnum	\count 2 < -32767
		\then	\divide \count 2 by 4
			\divide \count 4 by 4
		\else
		\fi
	\fi
	\multiply \count 0 by \count 2
	\divide \count 0 by \count 4
	\xdef \product {#1 = \the \count 0 \internal@nits}%
	\aftergroup \product
       }}

\def\r@duce{\ifdim\dimen0 > 90\r@dian \then   % sin(x) = sin(180-x)
		\multiply\dimen0 by -1
		\advance\dimen0 by 180\r@dian
		\r@duce
	    \else \ifdim\dimen0 < -90\r@dian \then  % sin(x) = sin(360+x)
		\advance\dimen0 by 360\r@dian
		\r@duce
		\fi
	    \fi}

\def\Sine#1%
       {{%
	\dimen 0 = #1 \r@dian
	\r@duce
	\ifdim\dimen0 = -90\r@dian \then
	   \dimen4 = -1\r@dian
	   \c@mputefalse
	\fi
	\ifdim\dimen0 = 90\r@dian \then
	   \dimen4 = 1\r@dian
	   \c@mputefalse
	\fi
	\ifdim\dimen0 = 0\r@dian \then
	   \dimen4 = 0\r@dian
	   \c@mputefalse
	\fi
	\ifc@mpute \then
		\divide\dimen0 by 180
		\dimen0=3.141592654\dimen0
		\dimen 2 = 3.1415926535897963\r@dian %%% a well-known constant
		\divide\dimen 2 by 2 %%% we only deal with -pi/2 : pi/2
		\Mess@ge {Sin: calculating Sin of \nodimen 0}%
		\count 0 = 1 %%% see power-series expansion for sine
		\dimen 2 = 1 \r@dian %%% ditto
		\dimen 4 = 0 \r@dian %%% ditto
		\loop
			\ifnum	\dimen 2 = 0 %%% then we've done
			\then	\stillc@nvergingfalse 
			\else	\stillc@nvergingtrue
			\fi
			\ifstillc@nverging %%% then calculate next term
			\then	\term {\count 0} {\dimen 0} {\dimen 2}%
				\advance \count 0 by 2
				\count 2 = \count 0
				\divide \count 2 by 2
				\ifodd	\count 2 %%% signs alternate
				\then	\advance \dimen 4 by \dimen 2
				\else	\advance \dimen 4 by -\dimen 2
				\fi
		\repeat
	\fi		
			\xdef \sine {\nodimen 4}%
       }}

\def\Cosine#1{\ifx\sine\UnDefined\edef\Savesine{\relax}\else
		             \edef\Savesine{\sine}\fi
	{\dimen0=#1\r@dian\multiply\dimen0 by -1
	 \advance\dimen0 by 90\r@dian
	 \Sine{\nodimen 0}
	 \xdef\cosine{\sine}
	 \xdef\sine{\Savesine}}}	      
\def\psdraft{
	\def\@psdraft{0}
}
\def\psfull{
	\def\@psdraft{100}
}

\psfull

\newif\if@draftbox
\def\psnodraftbox{
	\@draftboxfalse
}
\@draftboxtrue

\newif\if@prologfile
\newif\if@postlogfile
\def\pssilent{
	\@noisyfalse
}
\def\psnoisy{
	\@noisytrue
}
\psnoisy
\newif\if@bbllx
\newif\if@bblly
\newif\if@bburx
\newif\if@bbury
\newif\if@height
\newif\if@width
\newif\if@rheight
\newif\if@rwidth
\newif\if@angle
\newif\if@clip
\newif\if@verbose
\newif\if@scale
\def\@p@@sclip#1{\@cliptrue}

\def\@p@@sfile#1{\def\@p@sfile{null}%
	        \openin1=#1
		\ifeof1\closein1%
		       \openin1=\figurepath#1
			\ifeof1\typeout{Error, File #1 not found}
			   \if@bbllx\if@bblly\if@bburx\if@bbury% added 6/91 Rob Russell
			      \def\@p@sfile{#1}%
			   \fi\fi\fi\fi
			\else\closein1
			    \edef\@p@sfile{\figurepath#1}%
                        \fi%
		 \else\closein1%
		       \def\@p@sfile{#1}%
		 \fi}
\def\@p@@sfigure#1{\def\@p@sfile{null}%
	        \openin1=#1
		\ifeof1\closein1%
		       \openin1=\figurepath#1
			\ifeof1\typeout{Error, File #1 not found}
			   \if@bbllx\if@bblly\if@bburx\if@bbury% added 6/91 Rob Russell
			      \def\@p@sfile{#1}%
			   \fi\fi\fi\fi
			\else\closein1
			    \def\@p@sfile{\figurepath#1}%
                        \fi%
		 \else\closein1%
		       \def\@p@sfile{#1}%
		 \fi}

\def\@p@@sbbllx#1{
		\@bbllxtrue
		\dimen100=#1
		\edef\@p@sbbllx{\number\dimen100}
}
\def\@p@@sbblly#1{
		\@bbllytrue
		\dimen100=#1
		\edef\@p@sbblly{\number\dimen100}
}
\def\@p@@sbburx#1{
		\@bburxtrue
		\dimen100=#1
		\edef\@p@sbburx{\number\dimen100}
}
\def\@p@@sbbury#1{
		\@bburytrue
		\dimen100=#1
		\edef\@p@sbbury{\number\dimen100}
}
\def\@p@@sheight#1{
		\@heighttrue
		\dimen100=#1
   		\edef\@p@sheight{\number\dimen100}
}
\def\@p@@swidth#1{
		\@widthtrue
		\dimen100=#1
		\edef\@p@swidth{\number\dimen100}
}
\def\@p@@srheight#1{
		\@rheighttrue
		\dimen100=#1
		\edef\@p@srheight{\number\dimen100}
}
\def\@p@@srwidth#1{
		\@rwidthtrue
		\dimen100=#1
		\edef\@p@srwidth{\number\dimen100}
}
\def\@p@@sangle#1{
		\@angletrue
		\edef\@p@sangle{#1} %\number\dimen100}
}
\def\@p@@ssilent#1{ 
		\@verbosefalse
}
\def\@p@@sscale#1{
		\def\@p@scale{#1}
		\@scaletrue
}
\def\@p@@sprolog#1{\@prologfiletrue\def\@prologfileval{#1}}
\def\@p@@spostlog#1{\@postlogfiletrue\def\@postlogfileval{#1}}
\def\@cs@name#1{\csname #1\endcsname}
\def\@setparms#1=#2,{\@cs@name{@p@@s#1}{#2}}
\def\ps@init@parms{
		\@bbllxfalse \@bbllyfalse
		\@bburxfalse \@bburyfalse
		\@heightfalse \@widthfalse
		\@rheightfalse \@rwidthfalse
		\@scalefalse
		\def\@p@sbbllx{}\def\@p@sbblly{}
		\def\@p@sbburx{}\def\@p@sbbury{}
		\def\@p@sheight{}\def\@p@swidth{}
		\def\@p@srheight{}\def\@p@srwidth{}
		\def\@p@sangle{0}
		\def\@p@sfile{}
		\def\@p@scost{10}
		\def\@sc{}
		\@prologfilefalse
		\@postlogfilefalse
		\@clipfalse
		\if@noisy
			\@verbosetrue
		\else
			\@verbosefalse
		\fi
}
\def\parse@ps@parms#1{
	 	\@psdo\@psfiga:=#1\do
		   {\expandafter\@setparms\@psfiga,}}
\newif\ifno@bb
\def\bb@missing{
	\if@verbose{
		\typeout{psfig: searching \@p@sfile \space  for bounding box}
	}\fi
	\no@bbtrue
	\epsf@getbb{\@p@sfile}
        \ifno@bb \else \bb@cull\epsf@llx\epsf@lly\epsf@urx\epsf@ury\fi
}	
\def\bb@cull#1#2#3#4{
	\dimen100=#1 bp\edef\@p@sbbllx{\number\dimen100}
	\dimen100=#2 bp\edef\@p@sbblly{\number\dimen100}
	\dimen100=#3 bp\edef\@p@sbburx{\number\dimen100}
	\dimen100=#4 bp\edef\@p@sbbury{\number\dimen100}
	\no@bbfalse
}

\newdimen\p@intvaluex
\newdimen\p@intvaluey
\newdimen\@ffsetvalue
\newdimen\x@ffsetvalue
\newdimen\y@ffsetvalue

\def\compute@offset#1#2{{\dimen0=#1 sp\dimen1=#2 sp
			\advance\dimen1 by -\dimen0
			\dimen1=\sine\dimen1
			\dimen0=\cosine\dimen1
			\ifdim\dimen0<0sp \dimen1=0sp \fi
			\global\@ffsetvalue=\dimen1}}

\def\rotate@#1#2{{\dimen0=#1 sp\dimen1=#2 sp
		  \global\p@intvaluex=\cosine\dimen0
		  \dimen3=\sine\dimen1
		  \global\advance\p@intvaluex by -\dimen3
		  \global\p@intvaluey=\sine\dimen0
		  \dimen3=\cosine\dimen1
		  \global\advance\p@intvaluey by \dimen3
		  }}
\def\compute@bb{
		\no@bbfalse
		\if@bbllx \else \no@bbtrue \fi
		\if@bblly \else \no@bbtrue \fi
		\if@bburx \else \no@bbtrue \fi
		\if@bbury \else \no@bbtrue \fi
		\ifno@bb \bb@missing \fi
		\ifno@bb \typeout{FATAL ERROR: no bb supplied or found}
			\no-bb-error
		\fi
		\if@angle 
			\Sine{\@p@sangle}\Cosine{\@p@sangle}
			\compute@offset{\@p@sbblly}{\@p@sbbury}
			\x@ffsetvalue=\@ffsetvalue
			\compute@offset{\@p@sbburx}{\@p@sbbllx}
			\y@ffsetvalue=\@ffsetvalue

			\rotate@{\@p@sbbllx}{\@p@sbblly}
			\advance\p@intvaluex by -\x@ffsetvalue
			\advance\p@intvaluey by -\y@ffsetvalue
			\edef\@p@sbbllx{\number\p@intvaluex}
			\edef\@p@sbblly{\number\p@intvaluey}

			\rotate@{\@p@sbburx}{\@p@sbbury}
			\advance\p@intvaluex by \x@ffsetvalue
			\advance\p@intvaluey by \y@ffsetvalue
			\edef\@p@sbburx{\number\p@intvaluex}
			\edef\@p@sbbury{\number\p@intvaluey}
			{
			 \count0=\@p@sbbllx \count1=\@p@sbblly
		 	 \count2=\@p@sbburx \count3=\@p@sbbury
			 \dimen0=\@p@sbbllx sp\dimen1=\@p@sbblly sp
		 	 \dimen2=\@p@sbburx sp\dimen3=\@p@sbbury sp
			 \dimen203=\dimen2 \advance\dimen203 by -\dimen0
			 \dimen204=\dimen3 \advance\dimen204 by -\dimen1
			 \ifdim\dimen203<0sp 
			      \count203=\count2 \count2=\count0 
			      \count0=\count203 
			      \global\edef\@p@sbbllx{\number\count0}
			      \global\edef\@p@sbburx{\number\count2}
			 \fi
			 \ifdim\dimen204<0sp 
			       \count204=\count3
			       \count3=\count1
			       \count1=\count204
			       \global\edef\@p@sbblly{\number\count1}
			       \global\edef\@p@sbbury{\number\count3}
			 \fi
			}
		\fi
		\count203=\@p@sbburx
		\count204=\@p@sbbury
		\advance\count203 by -\@p@sbbllx
		\advance\count204 by -\@p@sbblly
		\edef\@bbw{\number\count203}
		\edef\@bbh{\number\count204}
}
\def\in@hundreds#1#2#3{\count240=#2 \count241=#3
		     \count100=\count240	% 100 is first digit #2/#3
		     \divide\count100 by \count241
		     \count101=\count100
		     \multiply\count101 by \count241
		     \advance\count240 by -\count101
		     \multiply\count240 by 10
		     \count101=\count240	%101 is second digit of #2/#3
		     \divide\count101 by \count241
		     \count102=\count101
		     \multiply\count102 by \count241
		     \advance\count240 by -\count102
		     \multiply\count240 by 10
		     \count102=\count240	% 102 is the third digit
		     \divide\count102 by \count241
		     \count200=#1\count205=0
		     \count201=\count200
			\multiply\count201 by \count100
		 	\advance\count205 by \count201
		     \count201=\count200
			\divide\count201 by 10
			\multiply\count201 by \count101
			\advance\count205 by \count201
		     \count201=\count200
			\divide\count201 by 100
			\multiply\count201 by \count102
			\advance\count205 by \count201
		     \edef\@result{\number\count205}
}
\def\@ScaleInHundreds#1{
		\in@hundreds{#1}{\@p@scale}{100}
		\edef#1{\@result}
}
\def\compute@wfromh{
		\in@hundreds{\@p@sheight}{\@bbw}{\@bbh}
		\edef\@p@swidth{\@result}
}
\def\compute@hfromw{
		\in@hundreds{\@p@swidth}{\@bbh}{\@bbw}
		\edef\@p@sheight{\@result}
}
\def\compute@handw{
		\if@height 
			\if@width
			\else
				\compute@wfromh
			\fi
		\else 
			\if@width
				\compute@hfromw
			\else
				\edef\@p@sheight{\@bbh}
				\edef\@p@swidth{\@bbw}
			\fi
		\fi
}
\def\compute@resv{
		\if@rheight \else \edef\@p@srheight{\@p@sheight} \fi
		\if@rwidth \else \edef\@p@srwidth{\@p@swidth} \fi
}
\def\compute@sizes{
	\compute@bb
	\compute@handw
	\compute@resv
}
\def\psfig#1{\vbox {
	\ps@init@parms
	\parse@ps@parms{#1}
	\compute@sizes
	\if@scale
                \if@verbose
                        \typeout{psfig: scaling by \@p@scale}
                \fi
                \@ScaleInHundreds{\@p@swidth}
                \@ScaleInHundreds{\@p@sheight}
                \@ScaleInHundreds{\@p@srwidth}
                \@ScaleInHundreds{\@p@srheight}
        \fi
	\ifnum\@p@scost<\@psdraft{
		\if@verbose{
			\typeout{psfig: including \@p@sfile \space }
		}\fi
		\special{ps::[begin] 	\@p@swidth \space \@p@sheight \space
				\@p@sbbllx \space \@p@sbblly \space
				\@p@sbburx \space \@p@sbbury \space
				startTexFig \space }
		\if@angle
			\special {ps:: \@p@sangle \space rotate \space} 
		\fi
		\if@clip{
			\if@verbose{
				\typeout{(clip)}
			}\fi
			\special{ps:: doclip \space }
		}\fi
		\if@prologfile
		    \special{ps: plotfile \@prologfileval \space } \fi
		\special{ps: plotfile \@p@sfile \space }
		\if@postlogfile
		    \special{ps: plotfile \@postlogfileval \space } \fi
		\special{ps::[end] endTexFig \space }
		\vbox to \@p@srheight true sp{
			\hbox to \@p@srwidth true sp{
				\hss
			}
		\vss
		}
	}\else{
		\if@draftbox{		
			\hbox{\fbox{\vbox to \@p@srheight true sp{
			\vss
			\hbox to \@p@srwidth true sp{ \hss \@p@sfile \hss }
			\vss
			}}}
		}\else{
			\vbox to \@p@srheight true sp{
			\vss
			\hbox to \@p@srwidth true sp{\hss}
			\vss
			}
		}\fi

	}\fi
}}
\def\psglobal{\typeout{psfig: PSGLOBAL is OBSOLETE; use psprint -m instead}}
\psfigRestoreAt

\newif\ifpdf
\ifx\pdfoutput\undefined
  \pdffalse
\else
  \pdfoutput=1
  \pdftrue
\fi

\ifpdf
  \usepackage[pdftex]{graphicx}
  \usepackage[pdftex]{color}
  \DeclareGraphicsExtensions{.pdf,.png,.jpg}
\else
  \usepackage[dvips]{graphicx}
  \usepackage[dvips]{color}
  \DeclareGraphicsExtensions{.eps,.epsi,.ps}
\fi

\usepackage{times}

\pagestyle{plain}

\def\midv{\mathop{\,|\,}}

\long\def\cbk#1{{\color{red}[CBK: #1]}}
\newlength\colwidth \setlength\colwidth{3.25in}

\title{The Partial Evaluation Approach to \\
Information Personalization}

\author{Naren Ramakrishnan and Saverio Perugini\\
Department of Computer Science\\
Virginia Tech, Blacksburg, VA 24061\\
Email:\{naren,sperugin\}@cs.vt.edu}

\begin{document}

\maketitle

\begin{abstract}
\noindent
Information personalization refers to the automatic adjustment of information content,
structure, and presentation tailored to an individual user. By reducing information
overload and customizing information access, personalization systems have emerged as an
important segment of the Internet economy. This paper presents a systematic modeling methodology  ---
PIPE (`Personalization is Partial Evaluation')  --- for personalization.
Personalization systems are designed and implemented in PIPE by modeling 
an information-seeking interaction in a programmatic representation. The representation
supports the description of information-seeking activities as partial information and their
subsequent realization by {\it partial
evaluation}, a technique for specializing programs. 
We describe the modeling methodology at a conceptual level and outline representational choices.
We present two application case studies that use PIPE for personalizing web sites and describe
how PIPE suggests a novel evaluation criterion for information system designs. Finally,
we mention several fundamental implications of adopting the PIPE model for personalization and
when it is (and is not) applicable.
\end{abstract}
\newpage
\tableofcontents
\newpage
%\begin{descit}{}
%
%\noindent
%I have observed that many good ideas start out by claiming too much territory for themselves,
%and eventually, when they have received their fair share of attention and respect, the air
%clears and it emerges that, though still grand, they are not quite so grand and all-encompassing
%as their proponents first thought. But that's all right. $\ldots$ That would be a fine start.
%\flushright{Douglas R. Hofstadter, in {\it Analogy as the Core of Cognition}}
%\end{descit}
\section{Introduction}

One of the main contributions of information systems research is
the development of models that allow the 
specification and realization of information-seeking activities.
Besides formalizing important operations, such
models provide a vocabulary with which to reason about the information-seeking activity. 
For instance, if an information space is modeled as a term-document matrix, then the 
vector-space model permits the view of retrieval as measuring similarities between document 
vectors. Similarly, the modeling of data as a set of relations in a database system
affords expressive
query languages such as SQL. Other models and modeling methodologies can be found
in interactive information retrieval applications~\cite{scatter-gather,tkde-navigation,rabbit}.
Our goal in this paper is to present a modeling methodology for information personalization.

Personalization constitutes the mechanisms and technologies required to
customize information access to the end-user.  It can be defined as
the automatic adjustment of information content, structure, and presentation
tailored to an individual user. The reader will be familiar with instances of
personalization such as web sites that welcome a returning user and 
recommender systems~\cite{adomavicius01expert-driven,specissue}
at sites such as {\tt amazon.com}. The scope of 
personalization today extends beyond web pages and web sites~\cite{terveen} to many different
forms of information content and 
delivery~\cite{cacm-broader,cacm-kantor,cacm-streams}.
The underlying algorithms and techniques range from simple keyword matching 
of consumer profiles, to explicit~\cite{ira,grouplens,phoaks} or 
implicit~\cite{cacm-jaideep,cacm-myra} capture of user interaction.

Despite its apparent popularity in reducing 
information overload on the Internet, personalization suffers from a lack of any rigorous model 
or modeling methodology. One of the main reasons is that there are `personal
views of personalization~\cite{cacm-personal}.' There are hence as many ways
to design and build a personalization system as there are interpretations for what
personalization means. Such a diversity presents a difficulty when studying conceptual
models of personalization, in general.

%Consequently, a large number of dichotomies have been proposed
%that classify personalization research according to the
%philosophies of the underlying domains, the proposers, and their
%parent communities. Thus, categorizations such as
%content-based versus collaborative~\cite{adomavicius01expert-driven,specissue},
%`customization versus transformation' \cite{adaptive-sites}, non-destructive
%versus destructive, `public transportation versus hot-rods' \cite{rus} have
%become widely accepted. These dichotomies are based on the form of personalization,
%the level at which it is targeted, and the types of information used in 
%delivering the personalization. 

We present the first 
(to the the best of our knowledge) 
systematic modeling methodology for information personalization.
Termed PIPE (`Personalization is Partial Evaluation')~\cite{naren-ic}, our methodology makes no commitments to a 
particular 
algorithm, format for information resources,
type of information-seeking activities or, more basically, the nature 
of personalization delivered. Instead, it emphasizes the modeling of an
information space in a way where descriptions of 
information-seeking activities can be
represented as partial information. Such 
partial information is then exploited (in the model) by
{\it partial evaluation}, a technique popular in the programming languages
community~\cite{jones}. 

%In important ways,
%PIPE advances personalization research as well as our understanding of how to build information
%systems. PIPE enables the view of personalization as
%specializing representations. 
%%personalization, not in terms of particular algorithms or targeting goals, but in
%%terms of representations (of information systems).
%%This viewpoint provides a framework to study questions 
%%such as `how should I design my information system so that it is personalized for my users?'
%It requires the modeling of {\it interaction} with an information
%system in a suitable representation. Once we conduct the modeling, personalization is achieved
%quite literally, `for free.' Our modeling methodology also allows
%personalization to be provided in conjunction with
%other information-seeking activities such as browsing.
%Finally, PIPE contributes a novel evaluation criterion 
%for information system designs. It relates
%personalization to the way an information system design is {\it factored}.
%The evaluation criterion allows us to qualify the {\it personability} of an
%information system for a particular information-seeking activity, for example
%`web site X is 60\% more personalized for activity Y than web site Z.'
%

While our ideas and results apply to many forms of computerized 
information systems (e.g., web-based, voice-activated),
we restrict our attention to web sites in this paper. 
Later in our discussion, we 
qualify the range of information systems technologies to which PIPE can be applied.

\subsection*{Reader's Guide}
Section~\ref{example} introduces the basic concepts of PIPE with the example of personalizing
a browsing hierarchy on the web. Section~\ref{basics} outlines the
PIPE modeling
methodology and how it can be used 
for representing a variety of situations. 
Section~\ref{studies} describes two
application studies that use PIPE for personalizing web sites. 
Evaluation aspects implied by PIPE as
a modeling methodology are also described here.
Section~\ref{discuss} 
describes connections between PIPE and other approaches, and carefully
qualifies situations where PIPE is
(and is not) applicable. 
%Emerging scenarios that can be realized with PIPE and PIPE-like methodologies are also presented here. 
Finally, Section~\ref{conclu} summarizes the major
contributions of this work.
\section{Motivating Example}
\label{example}

Consider a consumer visiting an automobile dealership
to purchase a vehicle. Here are two possible scenarios.

\newpage
\begin{descit}{Scenario 1}
\vspace{-0.1in}
\begin{tabbing}
{\bf Dealer:} \= Madam, are you looking to purchase a passenger vehicle?\\
{\bf Buyer:} \> Yes.\\
{\bf Dealer:} \> Do you have a particular manufacturer in mind?\\
{\bf Buyer:} \> I know that cars made by Honda have the highest safety approval rating.\\
{\bf Dealer:} \> That is true. Honda comes in seven colors. Do you have a preference for color?\\
{\bf Buyer:} \> The `cyclone blue' looks pleasing.\\
(conversation continues to ascertain further details of the vehicle)
\end{tabbing}
\end{descit}

\begin{descit}{Scenario 2}
\vspace{-0.1in}
\begin{tabbing}
{\bf Dealer:} \= Sir, may I interest you in anything?\\
{\bf Buyer:} \> I am looking for a sport utility vehicle.\\
{\bf Dealer:} \> Sure, do you have a particular manufacturer in mind?\\
{\bf Buyer:} \> Not really, but the vehicle should be Red and made in 2001.\\
{\bf Dealer:} \> I see. \\
{\bf Buyer:} \> And by the way, I don't care for the fancy doormats and fittings.\\
{\bf Dealer:} \> Of course.\\
(conversation continues)
\end{tabbing}
\end{descit}
\noindent
%The sophistication in these scenarios arises from the {\it mixed-initiative} nature of human
%conversations. 
In the first scenario, the conversation is directed by the dealer,
and the buyer merely answers questions posed by the dealer. The second scenario
resembles the first upto a point, after which the buyer takes the initiative and
provides answers `out of turn.' When queried about manufacturer, the buyer responds
with information about color and year of manufacture instead. Nevertheless,
the conversation is not stalled and both parties continue the dialog
to (eventually) complete the information assessment task. At each stage in the
above conversations, the buyer has the choice of proceeding along the lines of inquiry initiated
 by the dealer
or can shift gears and address a different aspect of information assessment. Scenarios that
`mix' these two modes of inquiry in such arbitrary ways constitute the 
scope of {\it mixed-initiative} interaction~\cite{mixed-notkin}.

\begin{figure}
\centering
\begin{tabular}{cc}
\includegraphics[width=8.25cm,height=6cm]{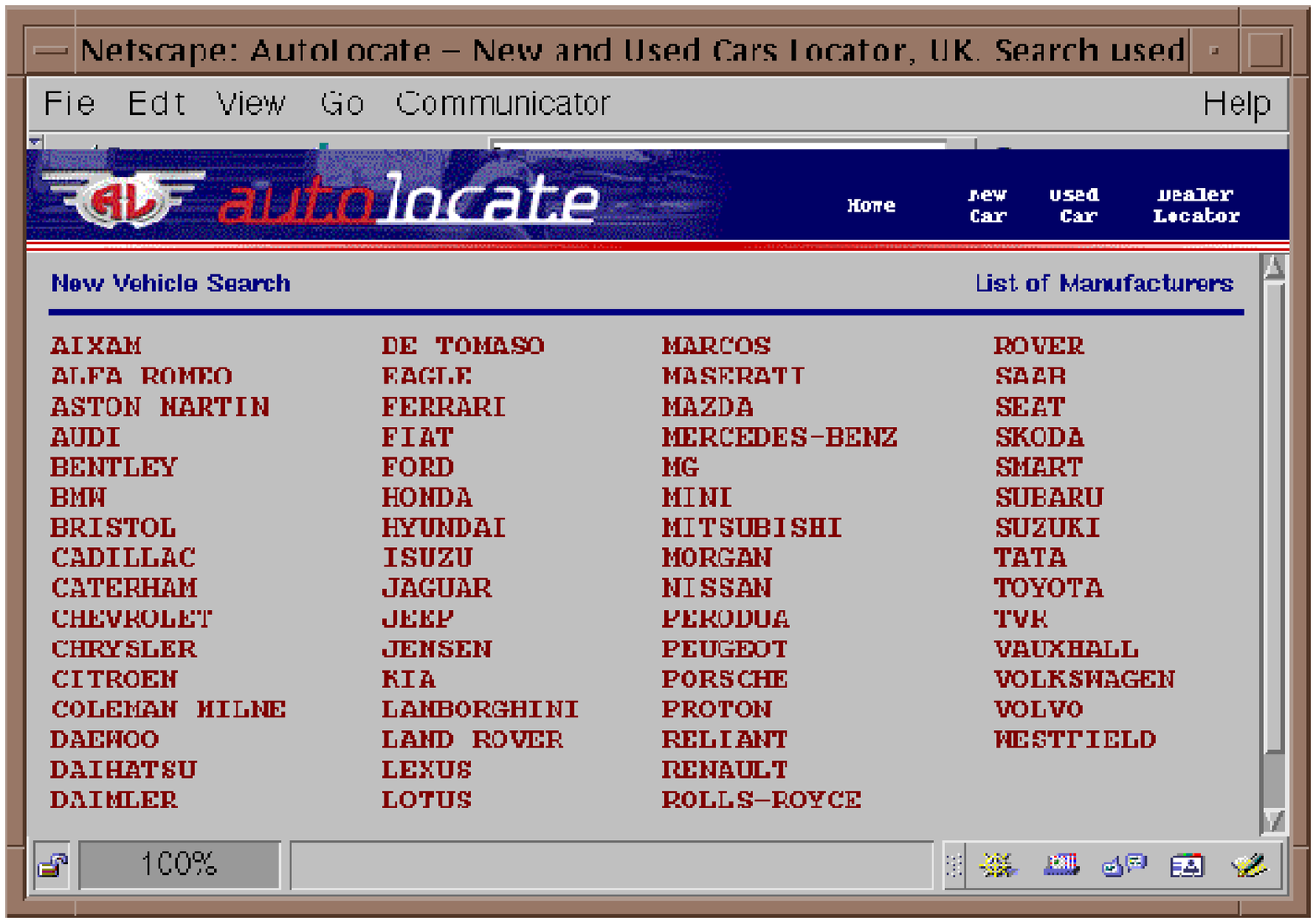}
\includegraphics[width=7.4cm,height=6cm]{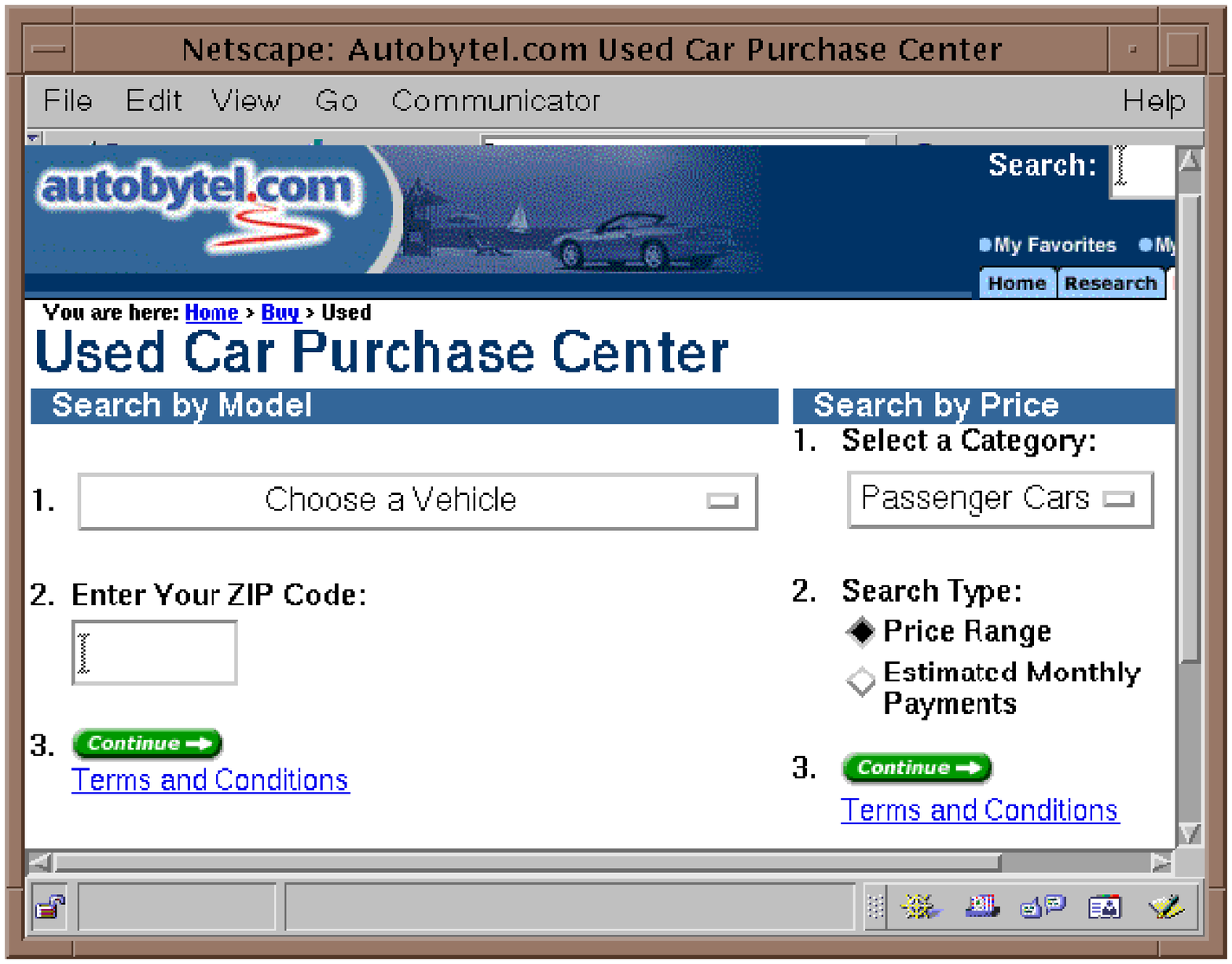}
\end{tabular}
\begin{tabular}{cc}
\includegraphics[width=8.25cm,height=9.89cm]{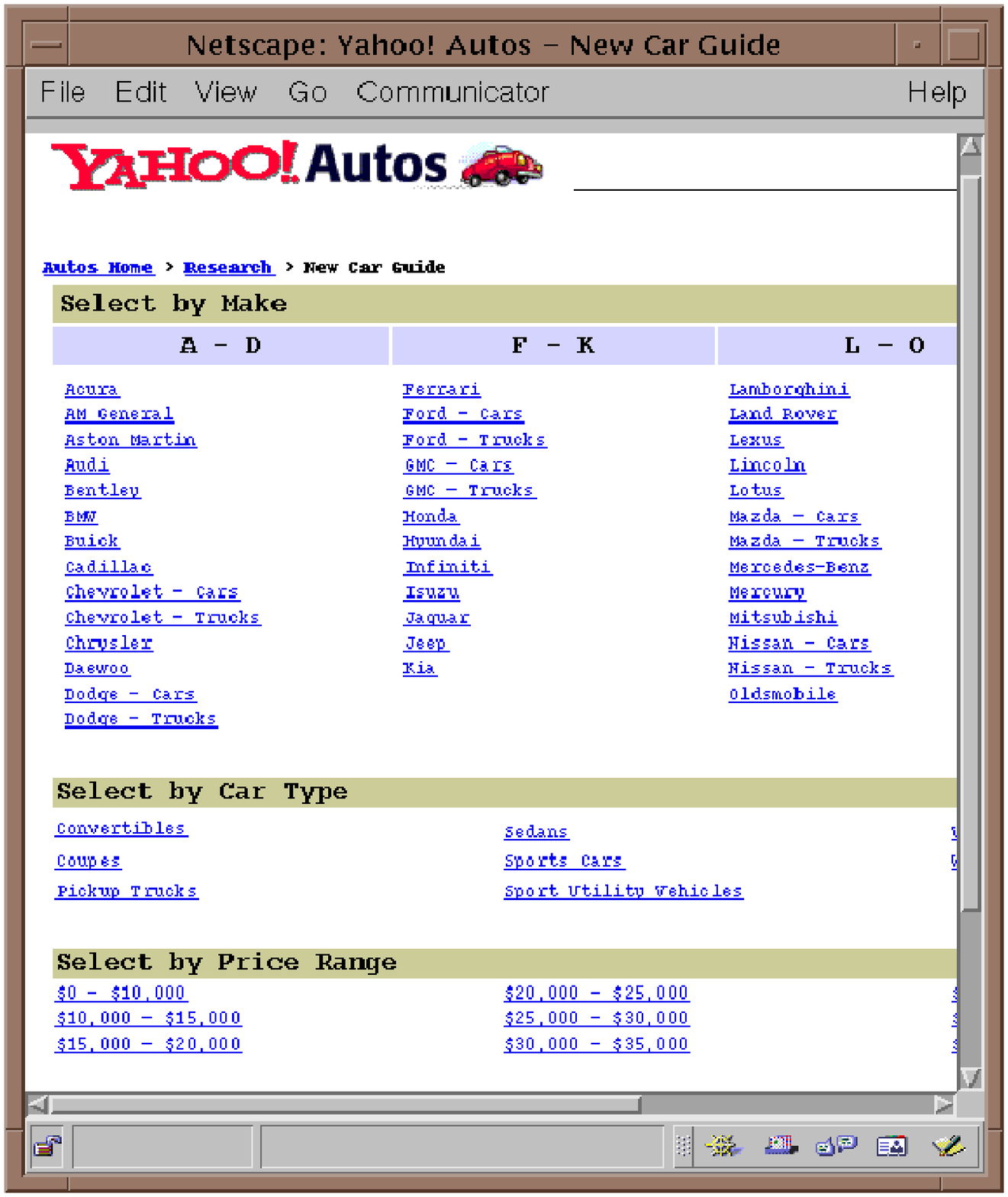}
\includegraphics[width=7.4cm,height=9.89cm]{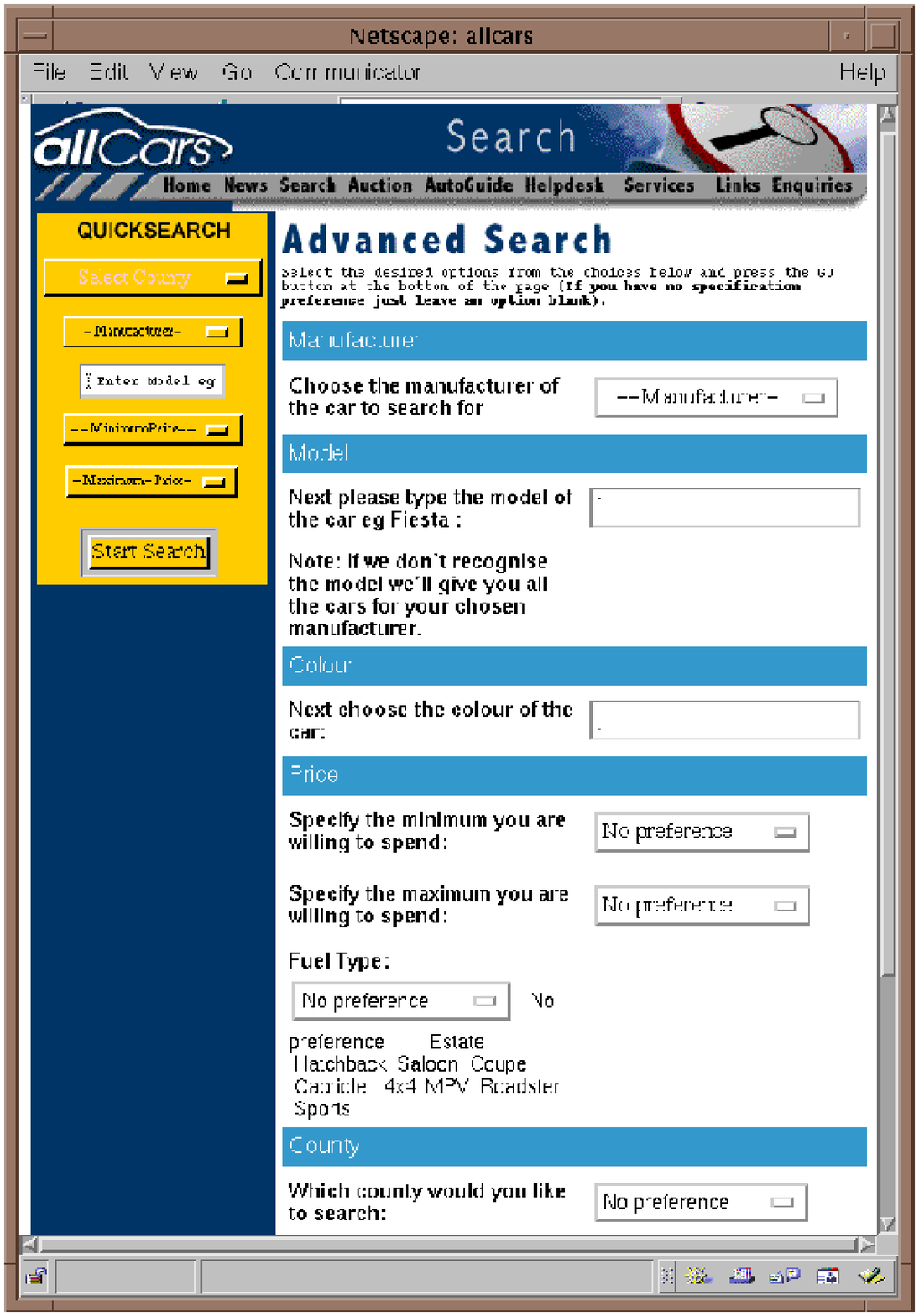}
\end{tabular}
%\begin{tabular}{cc}
%& \mbox{\psfig{figure=stupidautos.eps,width=5.5in}}
%\end{tabular}
\caption{Four typical solutions to organizing web catalogs. (top left)
A hardwired scenario.
(top right) A choice of two hardwired scenarios.
(bottom left) Complete enumeration involving all possible scenarios of interaction.
(bottom right) A `power-search' form that hides details of enumeration.}
\label{auto-solutions}
\end{figure}
Can we support a similar diversity of interaction in an online information system?
In other words, the system should have a default mode of interaction where a user
would fill in forms (or click on choices) in a specified order. A more
enterprising user should be able to supply any piece of information out 
of turn. Finally, it should be possible to
mix these two modes of interaction in any order. 
%While the term `mixed-initiative' affords many interpretations~\cite{mixed-notkin},
%we adopt an operational definition from our view of a dialog as
%an information assessment activity. By mixed-initiative, we emphasize the ability at any
%point to invoke (and later, return from) a subdialog whenever requested or specified by the
%user. This particular aspect of dialog structure in the context of information-seeking
%has been termed `goal-specification subdialogs' in~\cite{mixed-hci}.
%
At each stage of the interaction (whether system-initiated or user-requested), the system should
respond with the appropriate set of choices available. For instance, notice the 
restriction to
seven colors once the decision on Honda is made in {\it Scenario 1}. If the choice
of color was made at the outset, presumably more selections would have been available.
A system that supports such a diversity of interaction would be personalized to 
a user's individual preference(s) for information-seeking. 

The typical solution involves
anticipating the forms of interactions that have to be supported and designing interfaces
to support the implied scenarios (in this paper, we use the term `scenarios' to mean scenarios
of interaction).
Fig.~\ref{auto-solutions} describes four typical solutions that make various assumptions
on the scenarios that will be supported.
Fig.~\ref{auto-solutions} (top left)
can only support situations such as {\it Scenario 1} above, in that the user is forced to make 
a choice of manufacturer at the outset (and all remaining levels are similarly fixed).
We refer to this as a design that hardwires scenarios.
Fig.~\ref{auto-solutions} (top right) also hardwires scenarios,
but provides a choice of two such hardwired scenarios (i.e., search by model
or search by price). Fig.~\ref{auto-solutions}
(bottom left) is what we refer to as {\it complete enumeration}, which involves
enumerating all possible scenarios and providing interfaces to all of them~\cite{hearst-setting}.
While the interface in Fig.~\ref{auto-solutions} (bottom left) only depicts the top-level choice, we
could imagine that such multiplicity of choices are duplicated at all lower levels.
It is clear that enumeration could involve an exponential number of possibilities and correspondingly
cumbersome site designs.
And finally, Fig.~\ref{auto-solutions} (bottom right) provides the same
functionality as Fig.~\ref{auto-solutions} (bottom left)
but masks the details of enumeration in a convenient `power-search' form. 

All of these solutions rely on anticipating the points where an out-of-turn interaction
can occur and provide mechanisms to support it.  When opportunities 
for out-of-turn interaction are too restrictive, information systems cause major frustrations to users.
The basic problem is the representational mismatch between the user's mental model of the
information-seeking activity and the facilities that are available for describing the activity.

\begin{figure}
\centering
\begin{tabular}{cc}
\includegraphics[width=8.25cm,height=6.8cm]{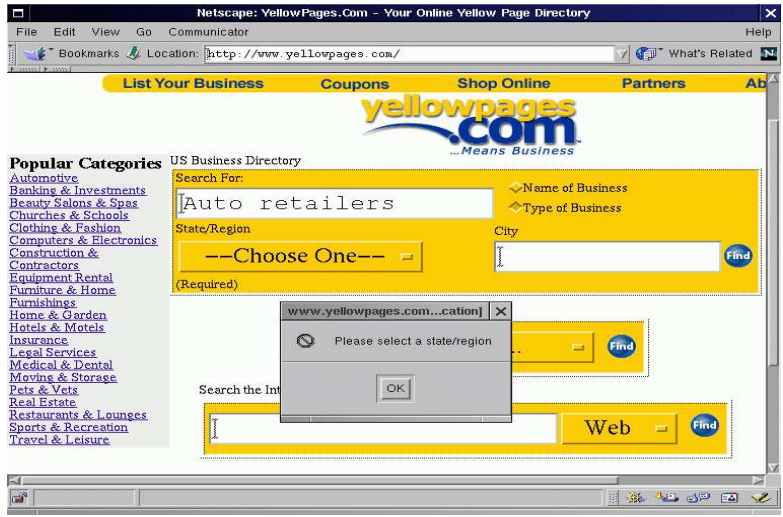}
%& \mbox{\psfig{figure=yellowstupid.eps,width=5.5in}}
\end{tabular}
\caption{An interface that prohibits certain information-seeking activities from
being decribed.}
\label{stupidyellow}
\end{figure}

In Fig.~\ref{stupidyellow}, the user is attempting to decide on an automotive retailer 
based on the services offered. He is open to
the possibility of traveling to a different city in order to make his purchase. He is thus
unsure of providing information about the location of the retailer, but the system insists that he
make this choice first. The reader can identify with examples such as these from other personal
experiences.

\subsection{The PIPE Approach}
We present an alternative design approach, one that promotes out-of-turn interaction without
predefining the points where such interaction can take place.
%and forms in which the information-seeker (buyer, in the above example)
%can barge in and alter the flow of interaction (conversation, in the above example). 
Consequently, the
interfaces produced by our approach are, at once, both more expressive and simpler than the ones 
in Fig.~\ref{auto-solutions}. 

\begin{figure}
\centering
\begin{tabular}{|l|l|} \hline
{\tt int pow(int base, int exponent) \{} & {\tt int pow2(int base) \{} \\
\,\,\,\,\,{\tt int prod = 1;} & \,\,\,\,\,{\tt return (base * base)} \\
\,\,\,\,\,{\tt for (int i=0;i<exponent;i++)} &  \} \\
\,\,\,\,\,\,\,\,\,\,{\tt prod = prod * base;} & \\
\,\,\,\,\,{\tt return (prod);} & \\
\} & \\
\hline
\end{tabular}
\caption{Illustration of the partial evaluation technique.
A general purpose {\tt pow}er function written in C (left) and
its specialized version (with {\tt exponent} statically set to 2) to handle squares
(right). Such specializations are performed automatically by partial evaluators such as C-Mix.}
\label{pe}
\end{figure}

\begin{figure}
\centering
\begin{tabular}{cc}
& \mbox{\psfig{figure=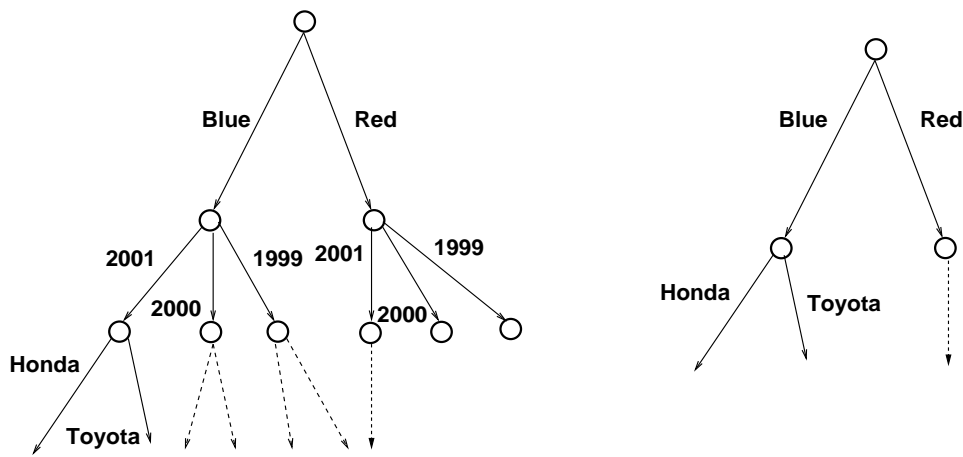,width=5in}}
\end{tabular}
\caption{Personalizing a browsing hierarchy. (left)
Original information resource. (right) Personalized hierarchy with respect to vehicles
made in 2001. Notice that not only the pages, but also their structure is customized for (further
browsing by) the user.}
\label{pipe-illustrate}
\end{figure}

\begin{figure}
\centering
\begin{tabular}{|l|l|} \hline
{\tt if (Blue)} & \\
\,\,\,\,{\tt if (2001)} & \\
\,\,\,\,\,\,\,\,{\tt if (Honda)} & \\
\,\,\,\,\,\,\,\,\,\,\,\,{$\cdots \cdots \cdots$} & {\tt if (Blue)}\\
\,\,\,\,\,\,\,\,{\tt else if (Toyota)} & \,\,\,\,{\tt if (Honda)} \\
\,\,\,\,\,\,\,\,\,\,\,\,{$\cdots \cdots \cdots$} & \,\,\,\,\,\,\,\,{$\cdots \cdots \cdots$}\\
\,\,\,\,{\tt else if (2000)} & \,\,\,\,{\tt else if (Toyota)}\\
\,\,\,\,\,\,\,\,{$\cdots \cdots \cdots$} & \,\,\,\,\,\,\,\,{$\cdots \cdots \cdots$} \\
{\tt else if (Red)} & {\tt else if(Red)} \\
\,\,\,\,{\tt if (2001)} &  \,\,\,\,{$\cdots \cdots \cdots$} \\
\,\,\,\,\,\,\,\,{$\cdots \cdots \cdots$} & \\
\,\,\,\,{\tt else if (2000)} & \\
\,\,\,\,\,\,\,\,{$\cdots \cdots \cdots$} & \\
\hline
\end{tabular}
\caption{Using partial evaluation for personalization. (left) Programmatic input to
partial evaluator, reflecting the organization of information in Fig.~\ref{pipe-illustrate} (left).
(right) Specialized program from the partial evaluator, used to create the personalized
information space shown in Fig.~\ref{pipe-illustrate} (right).}
\label{pe2}
\end{figure}

Let us begin by considering the scenario where a user obediently supplies information attributes
in the order requested. For ease of presentation, we assume that there are three attributes
--- color, year of manufacture, and manufacturer --- and that the information system ascertains
values for them in this order. The key contribution of PIPE is to cast this seemingly
inflexible and hardwired scenario in a representation that allows its automatic transformation
into other scenarios. In particular, PIPE represents an information space
as a program, partially evaluates the program with respect to (any) user input, and 
recreates a personalized information space from the specialized program. 

%\subsubsection*{Partial Evaluation}
The input to a partial evaluator
is a program and (some) static information about its arguments. Its
output is a specialized version of this program (typically in the same
language),
that uses the static information to `pre-compile' as many operations
as possible. A simple example is how the C function {\tt pow}
can be specialized to create a new function, say
{\tt pow2}, that computes the square of an integer.
 Consider for example,
the definition of a {\tt pow}er function shown in the left part of Fig.~\ref{pe}.
If we knew that a particular user will utilize it
only for computing squares of
integers, we could specialize it (for that user) to produce the {\tt pow2}
function.
Thus, {\tt pow2} is obtained automatically (not by a human programmer)
from {\tt pow} by precomputing all expressions that involve {\tt exponent},
unfolding the for-loop, and by various other compiler transformations such as
{\it copy propagation} and {\it forward substitution}.
Automatic program specializers are available for C, FORTRAN, PROLOG, LISP, and several other 
important
languages. The interested reader is referred to \cite{jones} for a good introduction.
While the traditional motivation for using partial evaluation is to achieve speedup
and/or remove interpretation overhead \cite{jones}, it can also be viewed as a technique
for simplifying program presentation, by removing inapplicable, unnecessary,
and `uninteresting' information (based on user criteria) from a program.

%\subsubsection*{Using Partial Evaluation for Personalization}
Consider the hardwired scenario depicted in Fig.~\ref{pipe-illustrate} (left).
We can abstract this hierarchy by the program
in Fig.~\ref{pe2} (left) whose structure models the information resource (in
this case, a hierarchy of web pages) and whose control-flow models 
the information-seeking activity within it (in 
this case, browsing through the hierarchy by making individual selections). The link
labels are represented as program variables and semantic dependencies between links
are captured by the mutually-exclusive {\tt if..else} dichotomies. As it is modeled in
Fig.~\ref{pe2} (left), the program reflects the assumption that
the choice of year is usually made at the second level, after a color selection has been
made. However, to personalize for the user who says `2001' at the outset, we partially
evaluate the program with respect to the variable {\tt 2001} (setting it to one and all
conflicting variables such as {\tt 2000} to zero). This produces the simplified
program in Fig.~\ref{pe2} (right), which can
be used to recreate web pages with personalized web content (shown in Fig.~\ref{pipe-illustrate},
right). The second level of the hierarchy is simplified,
bringing the originally third level as the new second level. The
user is able to provide the value of any deeply
nested variable out of turn, thus achieving mixed-initiative interaction.

%Executing the program in the
%order and form in which it was modeled amounts to the system-initiated mode of `browse as I say.'
%`Jumping ahead' to nested program segments by partially evaluating the program amounts to 
%the user-directed mode of personalization. 

\subsection{Some Preliminary Observations}
\label{observe}

Personalization systems are thus designed and implemented in PIPE by modeling
an information-seeking activity in a programmatic representation. The above
example has been carefully constructed to highlight the many advantages and
opportunities provided by PIPE. Before we describe PIPE in detail, it will be
helpful to summarize the lessons from the above example. 

\begin{enumerate}
\setlength{\leftmargin}{0in}
\item {\it PIPE equates personalization to specializing representations.}
%\vspace{-0.1in}
As a methodology, PIPE asserts that if interaction in an information space can
be represented as a program, then a personalized information space can be automatically
generated by partial evaluation. It is upto the designer to supply the representation
as a program and reinterpret the program in information systems terms.  
The meaning of the programmatic representation is thus
external to the basis for personalization (partial evaluation). 

For instance,
the act of clicking on the `Honda' hyperlink to browse through Honda cars is captured
in Fig.~\ref{pe2} by just the expression {\tt if (Honda)}. Clicking on the link amounts
to evaluating this conditional to be true. 
%A partial evaluator has no semantic 
%understanding of `clicking.' It is only the designer's creativity (or imagination) 
%that allows him to assign such an understanding. It is thus imperative that we 
%have a meaningful mapping 
%of programming constructs to aspects of
%interaction with an information system. 
%
%In some cases, these mappings are obvious. A 
The conditional construct {\tt if} is thus used as a logical
point where the state of
information is tested before proceeding any further. It could model either
a hyperlink that has to be clicked or a free-form text box whose entries are evaluated.
%Assigning an interpretation to constructs such as {\tt goto}
%is more tricky (but see Section~\ref{methods}).

\item {\it The effectiveness of PIPE depends on what is modeled (and how).}
The effectiveness of a PIPE implementation depends on the
the particular modeling choices made {\it within} the programmatic
representation (akin
to~\cite{rabbit}).
We cannot overemphasize this aspect --- the example
in Fig.~\ref{pe2} can be made `more personalized' by conducting
a more sophisticated modeling of the underlying domain. For instance, information such
as vehicle VIN numbers, history of ownership, 
mileage on the vehicle, and photos of the car can be further modeled as a browsable
hierarchy and `attached' (functionally invoked)  at various places
in the program of Fig.~\ref{pe2} (left). Conversely the example in Fig.~\ref{pe2} (left)
can be made `less personalized' by, for instance, requiring categorical information along with user input.
Replacing {\tt if (2001)} in Fig.~\ref{pe2} (left)
with {\tt if (Year=2001)} implies that the specification of the type of
input (namely that `2001' refers to the year of manufacture) is required
in order for the statement to be partially evaluated.
Personalization systems built with PIPE can thus be distinguished by what
they model and the forms of customization enabled by applying
partial evaluation to such a modeling.

Similarly, the way in which program variables are associated with user input 
can influence the effectiveness of a PIPE implementation.
Values for program variables 
could come from a content-based technique or a so-called
collaborative technique. For instance, the variable {\tt Honda} could be set to true, either
because the user explicitly said so, or because `Honda' was recommended to the user by an automatic
recommender system. In addition, different variables could afford different
interpretations.

Sometimes we can take advantage of a domain semantics when associating values
with program variables or in modeling the program.
Fig.~\ref{pe2} models a `strict' semantics of variable assignment 
by the {\tt if..else} dichotomies. If {\tt Blue} is evaluated to true, then every other option 
qualified by the {\tt else} constructs (such as {\tt Red}) would be automatically removed from 
further consideration. This is due to our assumption that if the user declares `Blue' as his
preference, then he would not be interested in Red cars. If such a semantics
is not appropriate, then we would not have {\tt else} clauses in our conditionals.
Thus, PIPE doesn't dictate what the domain semantics (for assigning program variables) should be 
or even that it should be available. But it can take advantage of a domain semantics, if one exists. 

%Consider again Fig.~\ref{pe2}, where the user
%declares `2001' as his preference and we partially evaluate
%w.r.t. {\tt 2001}. What is being simplified (actually, removed) in this example is
%the test {\tt if (2001)} and along with it, 
%a web page that requires the user to click on `2001.' 
%The reader might notice that we also set the program variable {\tt 2000} to be false. This
%is due to our understanding of how the personalization scenario is expressible as partial inputs.
%When the user declares `2001,' we can reason that she will not be interested in vehicles made
%in other years, and set the corresponding variables to zero. We reiterate that PIPE doesn't
%require that we set {\tt 2000} to be false; it comes from our understanding of what the user's
%interests are. 
%
%Likewise, if the user declares `Blue,' we can set the {\tt Blue} variable to one and all
%other conflicting variables such as {\tt Red} to zero. Once again, this doesn't mean that `Blue'
%implies `not Red'; it just means that we are using a domain semantics that enables us
%to make simplifications. If the user could still be interested in Red cars, then we
%do not set {\tt Red} to zero.
%
Finally, the translation of the program from and back to
the information space could be done in different ways. In Fig.~\ref{pe2} (left) we modeled the
program by abstracting hyperlinks across pages as conditionals. When we recreate personalized 
pages from
Fig.~\ref{pe2} (right) we are not obliged to this design choice. We could cascade all the
interactions to within a single page, for instance. PIPE only requires that the designer of
the information system has a way of going from an information space to a programmatic
representation, and back again. Section~\ref{basics} covers modeling options
in detail.
 
% how to set variables, how to employ a domain theory
% goal-oriented
% interaction sequences are bounded
% common criticism of our work is..
\item {\it PIPE separates modeling for a personalization system
from the operational aspect of personalization.}
Personalization systems are usually described in terms of the techniques
that provide personalization or the level at which the information is tailored. Due to the
variety possible, comparisons of personalization systems have been 
difficult to make.  PIPE, on the other hand, shifts the focus to modeling for 
a personalization system. Any form of personalization is possible if the modeled
program allows the pertinent scenarios to be expressible as partial inputs. In 
Fig.~\ref{pe2} 
we cannot personalize cars with respect to occupancy, not because of 
any fundamental limitation in our personalization methodology, but because 
{\tt occupancy} is not available as a program variable. Similarly, we cannot
personalize cars with respect to the {\it Edmund's Car Guide} recommendations, 
because the latter information resource has not been modeled.
The separation of modeling 
from the operational aspect of conducting personalization means that we can 
devote our attention to modeling the interaction in as sophisticated a manner 
as required. It also means that we have to distinguish between evaluating an
implementation of the PIPE
methodology from an evaluation of the methodology itself.

\begin{figure}
\centering
\begin{tabular}{cc}
\includegraphics[width=3in]{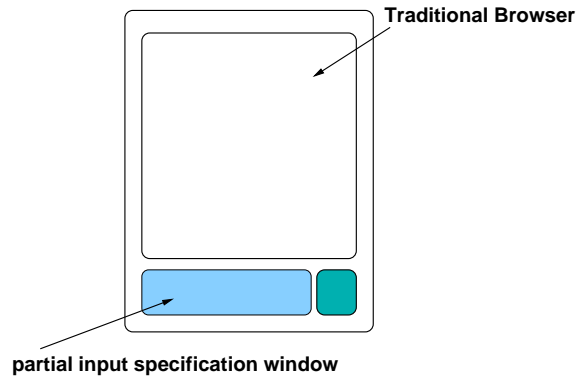}
%& \mbox{\psfig{figure=yellowstupid.eps,width=5.5in}}
\end{tabular}
\caption{Sketch of a PIPE interface to a traditional browser. The interface retains the
existing browsing functionality at all times. At any point in the 
interaction, in addition, the user has
the option of supplying personalization parameters and conducting personalization
(bottom two windows). Such an interface can be implemented as a toolbar option in existing
systems.}
\label{toolbar}
\end{figure}

%To our surprise, we have experienced remarkable resistance in propagating the view that
%PIPE is a modeling methodology, as different from a `system.' A frequent comment from visitors
%to our example applications at {\tt pipe.cs.vt.edu} is:
%\begin{descit}{}
%``PIPE is unimpressive since only subgraphs of the original site are being presented
%back to the user. No restructuring of the document content is provided.''
%\end{descit}
%\noindent
%This is a limitation of the {\it particular application} of PIPE (which does not model document
%content) than the methodology. Only
%the site graph (as in Fig.~\ref{pe2}) was modeled and hence only personalizations pertaining
%to manipulating the site graph are possible. 
%In Section~\ref{options} we outline a variety of ways to conduct more sophisticated modeling,
%including modeling document content. In Section~\ref{factor}, we address
%limitations of the PIPE modeling methodology itself.

% it is interaction that is personalized. not content-based collaborative, etc.
% it is upto us to interpret what the represntation means
% sometimes easy: if
% sammy: sometimes difficult: goto
% marcos: what is removed is the "link," We can go further.

% say dfs
% what is the representation: a compaction of interaction sequences
% define interaction sequence at this point
% once we know how to model interaction, we get P for free.
\item {\it The PIPE personalization operator is closed.}
Since the partial evaluation of a program results in another program, the PIPE
personalization operator is closed. In terms of interaction, this means that
any modes of information-seeking (such as browsing, in Fig.~\ref{pe2})
originally modeled in the program are preserved. In the above example, personalizing a browsable
hierarchy returns another browsable hierarchy.  The closure property also means that the
original information-seeking activity (browsing) and personalization can be interleaved in
any order. Executing the program in the
order and form in which it was modeled amounts to the system-initiated mode of `browse as I say.'
`Jumping ahead' to nested program segments by partially evaluating the program amounts to 
the user-directed mode of personalization. 
In Fig.~\ref{pe2}, the simplified program can be browsed in the traditional sense,
or partially evaluated further with additional user inputs. 
PIPE's use of partial evaluation is thus central to realizing a mixed-initiative mode of 
information-seeking, without explicitly hardwiring all possible scenarios 
of interaction (including out-of-turn interactions). A sketch of an interface design for
such mixed-initiative interaction is provided in Fig.~\ref{toolbar}.

\item {\it PIPE is most advantageous in information spaces that afford nested representations
of interactions and where information-seeking activities can involve out-of-turn interactions.}
For browsing hierarchies, a nested programmatic model can be trivially built by a depth-first 
crawl of the site (as in Fig.~\ref{pe2}). Not only is this modeling appropriate, it is
also concise and makes the advantages of partial evaluation obvious. 

\begin{figure}
\centering
\begin{tabular}{|l|} \hline 
{\tt if (Returning Customer)} \\
\,\,\,\,{\tt /* be nice to her */} \\
{\tt else} \\
\,\,\,\,{\tt /* just show usual catalog */} \\
\hline
\end{tabular}
\caption{A modeling of an information space that involves only one level of interaction.}
\label{pe4}
\end{figure} 

%Even though the hierarchy is designed in a color-year-model order, PIPE
%can accept values for program variables in any order.
%Further, the size of the program in Fig.~\ref{pe2}
%(left) is a small fraction of what it would be if we were to enumerate scenarios in an exhaustive
%fashion. Recall that enumerating scenarios to support all forms of barge in leads to
%cumbersome designs such as Fig.~\ref{auto-solutions} (bottom left).
%In addition to mirroring the browsing hierarchy, our nested representation makes the advantages 
%of partial evaluation obvious. 
%
On the other hand,
consider a web site 
that determines (perhaps by a cookie~\cite{cookies}) if a user is a 
returning customer and does something different based on this information. 
Modeling (only this) interaction can be done by the program in Fig.~\ref{pe4}. 
While partial evaluation is still applicable, it cannot do anything fancy since 
there is only one variable ({\tt Returning Customer}) to specify values for. There is
no deeply nested variable whose value can be supplied out of turn.
%In addition
%A nested representation is possible only if the space of information-seeking 
%activities is structured for us to take advantage of commonality across different
%scenarios of interaction.

Similarly, if all users would like to browse through the catalog in Fig.~\ref{pe2} by a 
color-year-model motif, then there is really only one way in which the catalog is being used.
This usage mirrors the way in which the catalog is modeled, without any out-of-turn interactions.
Partial evaluation is thus not necessary to support the information-seeking goals of any user. 

The presence of out-of-turn interactions implies different rates of specification for 
different aspects of information seeking, causing a rich variety of possible interactions.
In such a case, PIPE can be viewed as a technique that realizes a particular interaction sequence 
by combinations of simplification and normal execution.
In Section~\ref{factor}, we show more formally which representations (and 
which information spaces) are best suited for personalization by partial evaluation.

\end{enumerate}
\section{Essential Aspects of PIPE}
\label{basics}
We now describe the PIPE methodology in more detail and outline choices available for modeling
typical situations.
While partial evaluation permits formal specification with mathematical notation~\cite{jones-book}, we do not
take this approach here. Instead, for the {\it ACM TOIS} audience,
we aim to emphasize the larger context in which partial evaluation is used  in PIPE
and describe its advantages for information systems. We intend to present the formal aspects 
of the PIPE methodology in a second paper.

\subsection{Modeling Methodology}
\label{methods}

As a modeling methodology, PIPE only makes the weak assumption that information is
organized along a motif of interaction sequences. For our purposes, an interaction sequence is 
a list of primitive inputs used to describe the information-seeking
activity.  For instance in Fig.~\ref{pe2}, information about vehicles is organized along 
a color-year-model motif with the primitive inputs corresponding to specific choices 
of color, year, or model. The interaction sequence in this example involves the 
the choice of {\tt 2001} for {\tt year}, in support of the user's goals.

Information is embodied in an interaction sequence in two forms --- {\it structural}
and {\it terminal}. Structural information is what helps us refer to an interaction sequence;
it is explicitly represented in PIPE and specified via program variables. 
In Fig.~\ref{pe2}, the structural information corresponds to choices of color, year, and model.
This form of information thus captures the partial information supplied by the user
by instantiating parts of the motif. When the user specifies `2001' in Fig.~\ref{pe2}, the
{\tt year} part of the motif is turned on and set to this value. 

Terminal information is also represented in PIPE, but is not directly manipulatable or even
directly addressable. Programs in PIPE are not explicitly parameterized by 
this information and so the user cannot specify personalization in these terms. 
In Fig.~\ref{pe2}, terminal information corresponds to the leaves, which would be information 
about particular vehicles. In a different application, terminal information could reside at 
every step in the interaction sequence. 

Structural information provides the `backbone' that strings together terminal information.
However, it is important to note that
structural information is considered first-class information in PIPE and not merely `features' with 
which we index the `real information' (although it is tempting to view it this way).
To see why, observe that partial evaluation does not provide a mapping
from structural to terminal information (unless it was a complete evaluation specifying all
program variables).
After a partial evaluation (e.g., Fig.~\ref{pe2} (right)) the specialized program might still
contain structural information. This does not necessarily mean
that the user's information-seeking activity is incomplete. The residual structural 
information contributes to the programmatic modeling of interaction, {\it which is} the 
personalized information space in PIPE. Another way to see this is to note
that PIPE simplifies {\it interaction} with an information space. Thus interaction
can be seen to be the determiner of information (both structural and terminal). The view
of structural information as first-class information is also natural if we think of the program
in logic programming terms, rather than imperative programming.

%The issue of whether structural information is first-class is actually very fundamental to
%information system design. In Section~\ref{factor} we show the widespread ramifications of this idea.

Since information can be organized all along the interaction sequence, in both structural and terminal
forms, we need a way to define the state of information described by the sequence as a whole.
It is useful to assume a `combining function' for defining the state of 
information at the end of the sequence. A simple example of a combining function is the additive
operator which mirrors the accumulation of information by following an interaction sequence.
In Fig.~\ref{pe2}, if the color and model parts of the motif are turned on, then the state
of information known about that sequence is a set of values for \{{\tt color,model}\}.
Another example is to just retain information from the most recent step(s) in the sequence. This
would be appropriate when information-seeking has an exploratory nature to it and we wish to
discount some earlier steps in an interaction sequence as being `tentative' (the applications
presented in this paper do not have this flavor). Combining functions for terminal information
can be defined similarly.

Since PIPE only emphasizes the design and implementation of personalization systems, it doesn't
pay any attention to how the interaction sequences are obtained and how the choice between
terminal and structural parts is made. In particular, PIPE is not a 
complete lifecycle model for personalization system design and doesn't address issues such
as requirements gathering. Interaction sequences could come from explaining users' 
behavior~\cite{pipe-tochi,footprints}, by identifying all possible paths through a given
site, or from our conceptual understanding of the information-seeking activity. They also depend
on the targeting goals of the personalization system. In~\cite{pipe-tochi}, we have
presented a systematic methodology for obtaining interaction sequences and identifying
structural and terminal parts, by `operationalizing' scenarios of interaction; we refer the
reader to this reference for details. In this paper, we assume that they are available
and proceed to further characterize and represent them.

\subsubsection*{Characterizing Interaction Sequences}
Information seekers forage in different ways~\cite{pirolli-chapter} and the existing
design of the information system also influences their interaction sequences. 
An important aspect of an interaction sequence is its length, which affects
its subsequent representation in PIPE.

In many applications, interaction sequences are bounded. For instance,
in Fig.~\ref{pe2} an interaction sequence of length at most 3 describes
the information-seeking activity. Such sites and applications are characterized by their support
for a goal-oriented, opportunistic view of information-seeking. Hierarchies, recommender systems,
and scrolling to a specific location on a page are examples. 
In general, any information-seeking
activity that has clear start and end states and which relies on perceptual, display-driven
 clues that focus attention can be represented as a bounded sequence.

In other important cases, interaction sequences can be unbounded. The trivial example is when 
we allow the possibility that a user
may click `back buttons.' If we undo these steps before representation, we can proceed as if
they never happened. Alternatively, we can model back buttons using a finite-state machine (FSM), 
but we have to find a characterization of applications where modeling at this level 
of detail would be useful. A more interesting example of unbounded sequences involves browsing at a
site based on social network navigation, such as
{\tt www.\hskip0ex imdb.\hskip0ex com}. There are no leaves in this site and the site graph
resembles a social network.
Users are encouraged to systematically explore relationships between actors, movies, and directors
by `jumping connections.' Such
a site is characterized by an exploratory nature of information-seeking, akin to data mining. Goals are
articulated less clearly and cognitive knowledge is used from various resources to decide on how to
conduct information-seeking. In fact, there is no distinction between structural and terminal
information in this site! Any particular web page could be used to address other items or thought
of as the result of an information-seeking activity.

Both bounded and unbounded interaction sequences can be described using constructs such 
as regular expressions, grammars, FSMs, and programs; unbounded interaction sequences require 
special handling, due to the reasons mentioned above. In this paper, we concentrate on personalization 
applications describable by bounded interaction sequences and which have a clear separation between
structural and terminal parts.

\subsubsection*{Representing Interaction Sequences in PIPE}
Given that we can represent information-seeking activities as interaction sequences,
the set of scenarios that are likely to be encountered (over all users, perhaps)
can be represented by a corresponding set of interaction sequences. Representing this latter set 
faithfully and compactly as a program is key to the application of PIPE.  Once again, PIPE doesn't
indicate what this set should be: whether it is across all users~\cite{footprints}, whether it is for a 
group of users~\cite{adaptive-sites}, or whether it comes from our conceptual understanding 
of information-seeking.

For instance, Fig.~\ref{pe2} uses a nested representation to form the program for subsequent 
partial evaluation. Not only does it model the color-year-model motif (as it would have
been observed), it also allows us to model the year-color-model motif (by one 
partial evaluation). Since PIPE provides out-of-turn personalization, it is not necessary to
represent every interaction sequence explicitly in the program.

Compaction of interaction sequences is important for two reasons. The first is that it
 preserves the inherent structure of the (unpersonalized) information-seeking activity (such as browsing,
in Fig.~\ref{pe2}). This is useful in realizing mixed-initiative interaction with PIPE. Another reason is that
compaction permits scalable personalization solutions. 

Structural parts of interaction sequences can be represented using constructs in
a full-fledged programming language, such as C (as done in Fig.~\ref{pe2}) or LISP. 
A programming language provides many facilities that can help in compaction of interaction sequences. For example,
if we notice that all interaction sequences at a site require registration at some point in the interaction,
then the steps associated with registration could be factored out and procedurally invoked from various
other locations.
Off-the-shelf
partial evaluators (such as C-Mix) can then be used for specializing the representations.

It is important that we also model terminal parts of interaction sequences. In the example
of Fig.~\ref{pe2}, if there is text anchoring every hyperlink, then we can define
a program variable to start accumulating text once every conditional is evaluated to be true. This
could be achieved using associate arrays or by dynamic memory allocation constructs (e.g., pointers).
After partial evaluation, we can inspect the contents of this data structure at every stage
to present personalized (terminal) content. Inspecting the contents of the sequence as a whole will
provide an overall summary of the terminal information. Inspecting the contents of subsequences 
will provide
more fine-grain summaries of terminal information.

\begin{figure}
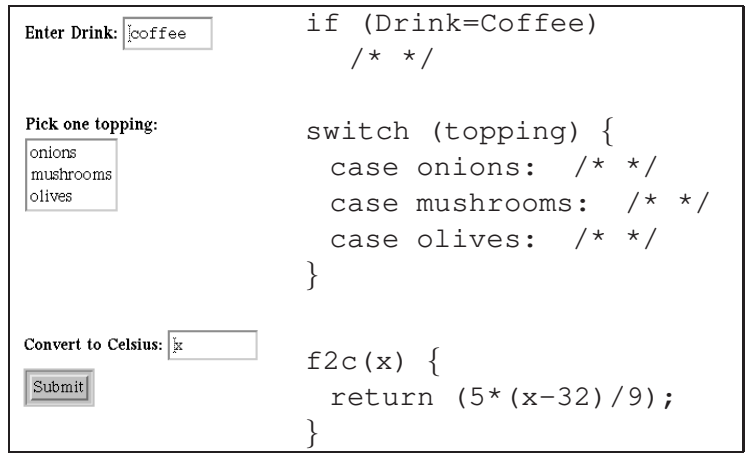

\centering
\begin{tabular}{|ll|}
\hline
\includegraphics[width=2.5cm,height=0.45cm]{display3.epsi} &
\begin{tabular}{l}
{\tt if (Drink=Coffee)}\\
\,\,\,\,\,\,\,\,\,{\tt /* */}
\end{tabular}\\
 & \\
\includegraphics[width=1.75cm,height=1.25cm]{display2.epsi} & 
\begin{tabular}{l}
{\tt switch (topping) \{}\\
\,\,\,\,\,{\tt case onions: /* */}\\
\,\,\,\,\,{\tt case mushrooms: /* */}\\
\,\,\,\,\,{\tt case olives: /* */}\\
{\tt \}}
\end{tabular} \\
 & \\
\includegraphics[width=3.1cm,height=1cm]{display4.epsi} &
\begin{tabular}{l}
{\tt f2c(x) \{}\\
\,\,\,\,\,{\tt return (5*(x-32)/9);}\\
{\tt \}}
\end{tabular} \\ \hline
\end{tabular}

\caption{Choices for representing aspects of interaction in PIPE.}
\label{cute-diagram}
\end{figure}

\subsubsection*{Creating a Personalization System}
To effect the creation of a personalization system, we define ways for the user to
specify values for program variables and a procedure by which personalized information
content is presented back to the user. Every construct used in the programmatic modeling (terminal or
structural) should be translatable into information systems terms, and vice versa.

Typically, there is a one-one mapping between interactions and programming
constructs. In Fig.~\ref{cute-diagram},
the textbox corresponds to a conditional, the listbox to a {\tt switch} construct, and
the unit convertor to a function in a PIPE modeling. 

Such mappings have to be revisited after partial evaluation.
For instance, the {\tt if} construct in Fig.~\ref{cute-diagram}
will either be removed or left as-is by a partial evaluation. This will just correspond
to removing or retaining the textbox in the personalized web site.  The {\tt switch} construct
in Fig.~\ref{cute-diagram} corresponding to a listbox is more interesting. After partial evaluation,
it might be the case that only one of the three topping options are left. Perhaps the person is allergic
to mushrooms and olives and we set those variables to zero. In this case, the
partial evaluator might remove the {\tt switch} altogether and replace it with a simple
{\tt if}. We can view this as a hint to render the listbox as a hyperlink in the personalized site.
Finally, the unit conversion utility in Fig.~\ref{cute-diagram} can be modeled in several ways.
We can view it as a functional black-box and model in PIPE the act of getting a value and
passing it to, say, a server-side script that performs the conversion. If we take this approach, we
should ensure that partial evaluation either retains the black-box representation or removes it;
it shouldn't `open' it up. Alternatively, we can explicitly
open up this black-box and model its contents as a function in a PIPE modeling (as done in Fig.~\ref{cute-diagram}).
As a functional modeling, PIPE thus enables the view of information systems as transducers.

In some cases partial evaluators, by their sophisticated support for program specialization,
cause difficulties. For instance, 
the technique of {\it program-point specialization}~\cite{jones} introduces copies
of functions at various places in the specialized program, tailored to specific situations. In
information systems terms, this amounts to creating content (structural as
well as terminal) that didn't exist before. In such a
case, we need to carefully interpret the meaning of the specialized representation.

Another caveat is that partial evaluation can sometimes induce {\tt goto}s in the specialized
program. We can view {\tt goto}s as
suggesting means by which the site design could be structured. If there is a {\tt goto}
from a point $A$ in the program to another point $B$, it just means that the information system
corresponding to point $B$ can be arrived at in many ways via interaction sequences and hence is
advantageous if factored out.

Finally, a semantics of values for program variables has to be defined.
In partial evaluation, values may be either specified or left unspecified.
By default, variable values cannot be weighted unless explicitly 
modeled in the PIPE program. However, techniques
such as query expansion can be employed to obtain values for other program variables. For instance, if
a user says `Honda' and a PIPE program models Honda cars under `Japanese automakers,'
then we can turn both these variables on
for the purposes of personalization. Semantics for program variables can also
be defined to take advantage of other taxonomical relationships in hierarchies~\cite{taxo-assign}.

\subsubsection*{A Salient Feature of PIPE}
An important advantage of PIPE is that while we provide options for modeling, there is
is no explicit step for describing how to implement personalization. Due 
to the sophistication of our representation, personalization will be
achieved if program variables (which correspond
to structural information) are available for partial evaluation.
This is in contrast to other modeling methodologies~\cite{statecharts,autoweb,rus} where personalization has to be provided 
as an explicit function from the conceptual design stage.

% 1. Modeling site structure: dfs, site graph, from navigational schema (see program compaction below)
% 2. information integration: composing subsystems, e.g., multiple sites. 
%To do this effectively, we need wrappers (for handling syn. poly. problems)
% 3. crawling clickable maps:
% 4. interacting with recommender systems: rec. sys. could be a f() within the model. Or, the aspect of recommendation could
%be "external" to PIPE, in that collaborative features correspond to program variables. "x" means. 
% 5. browsing computed information: hooks, 
% 6. modeling within a web page: virtual nodes, DTDs, naveen ashish, XTRACT, 
% 7. program compaction: put 4 pictures

%senators: 1,6
%gams: 1,2,4,6,7
%pigments: 1,2,4,5,6

% generators of hierarchies (the one you found with Sammy)
\subsection{Representational Choices}
\label{options}

Our primary example of modeling thus far addressed
navigation down a hierarchy via nested conditionals (see Fig.~\ref{pe2}).
This is one of the most common sources of bounded sequences; it can be obtained either
by explicit crawling or as graph representations of site structure from website 
management tools~\cite{webSiteManagement,webstrudel}. In the former,
extra care should be used to address purely navigational links (like a `Go Back'
button) and irregularities in web page authoring. Representations obtained from the latter
case are more robust since they directly enable the modeling of interaction sequences in terms of
directed labeled graphs~\cite{dataontheweb} or web schema~\cite{autoweb}. 

In this section, we present a number of other modeling options for personalization 
applications described by bounded interaction sequences. 

\subsubsection*{Interacting with Recommender Systems}
A recommender system can be viewed in PIPE as a way to set values for program variables or as a
function to be modeled. In the first case, the recommender is abstracted as a black-box and is external
to the program. Consider a recommender system at a third-party site that suggests 
automobile dealers based on 
experiences of its users. In such a case, we can invoke the facility to obtain values for program variables
which are then subsequently used for personalization. Alternatively, the functioning of the recommender 
can be explicitly modeled in PIPE. This allows the possibility that even its operation could 
be personalized. For instance, if the recommender system can suggest dealers all across the United States,
we can personalize its operation to only recommend dealers in a particular geographical region. This will
not be possible in the black-box modeling unless the recommender allows such explicit specification.

\subsubsection*{Information Integration}
Effective personalization scenarios require the integration of information from multiple sites.
Consider personalizing stock quotes for potential investors. The Yahoo!
Finance Cross-Index at {\tt quote.yahoo.com} provides a ticker symbol lookup for stock charts, 
financial statistics, and links to company profiles. It is easy to model and personalize
this site by the methods described above. 
However, what if the user desires to browse this site based on recommendations from an 
online brokerage? Besides support for cascading information flows, care should be taken
to ensure that structural information across multiple sites is correctly
cross-referenced. The online brokerage might refer to its 
recommendations by company name (e.g., `Microsoft'), while the Yahoo! cross-index 
uses the ticker symbol (`MSFT'). Standard solutions based on wrappers~\cite{kush-ai} and mediators can
be employed here~\cite{db-www,Ariadne}.
In PIPE, the individual interaction sequences from multiple sites can be cascaded in 
sequence to provide support for such integration scenarios, as shown in Fig.~\ref{ii}.

\begin{figure}
\centering
\begin{tabular}{|l|} \hline
{\tt main() \{} \\
\,\,\,\,{\tt /* invoke online brokerage */} \\
\,\,\,\,{\tt /* transforms from company name to ticker symbol */} \\
\,\,\,\,{\tt /* modeling of yahoo cross-index */} \\
{\tt \}} \\
\hline
\end{tabular}
\caption{Modeling information integration in PIPE.}
\label{ii}
\end{figure}

\subsubsection*{Modeling Clickable Maps}
Many web sites provide clickable image maps (e.g., JAVA/GIF) as 
interfaces to information. This is especially true
for weather sites, bioinformatics resources, and sites that involve modeling spatial information. Interpretation
is attached to clicking on particular locations of the map (for instance, `click on the state for which
you would like the weather').
Using data mining techniques~\cite{fukuda} and 
by sampling clicks on the map (and determining which pages they lead to), we can functionally model 
a clickable map in PIPE to arrive 
at constructs such as: `Choosing Wyoming on the United States map corresponds to clicking 
within $[a,b] \times [c,d]$.' Non-rectangular areas are described by unions of isothetic regions by
the data-mining technique described in~\cite{fukuda}.
Given such a representation, partial evaluation can remove portions of 
the image map based on user preferences. At this stage, we can reconstruct a personalized clickable
map by reversing the mapping or use attributes such as color and shade to highlight the selected
regions (for instance, to show only those regions on the map where air travel is delayed). We can also represent 
the personalized information in non-graphical terms. This option is useful not just for personalization 
but for improving the accessibility of information systems. A mobile handheld device incapable 
of presenting graphical content can take advantage of such modeling.

\subsubsection*{Modeling within a Page}
In some cases, it is necessary to model interaction sequences within a web page.
For instance, if a user is eyeballing a web page to look for telephone numbers of
an individual, then modeling the web page at this level of granularity and providing a program variable
for telephone number would be useful. Algorithms for mining structure within a 
web page (e.g., DTDs) ~\cite{naveen,craven-aij,xtract}
and for document segmentation~\cite{rus} can be used to arrive at compact 
representations of within-page interaction sequences. This
provides a richer set of features with which to conduct personalization. For instance,
partial evaluation can be used to remove complete sections of documents (e.g., intrusive
advertisement banners) when rendering the personalization. 

\begin{figure}
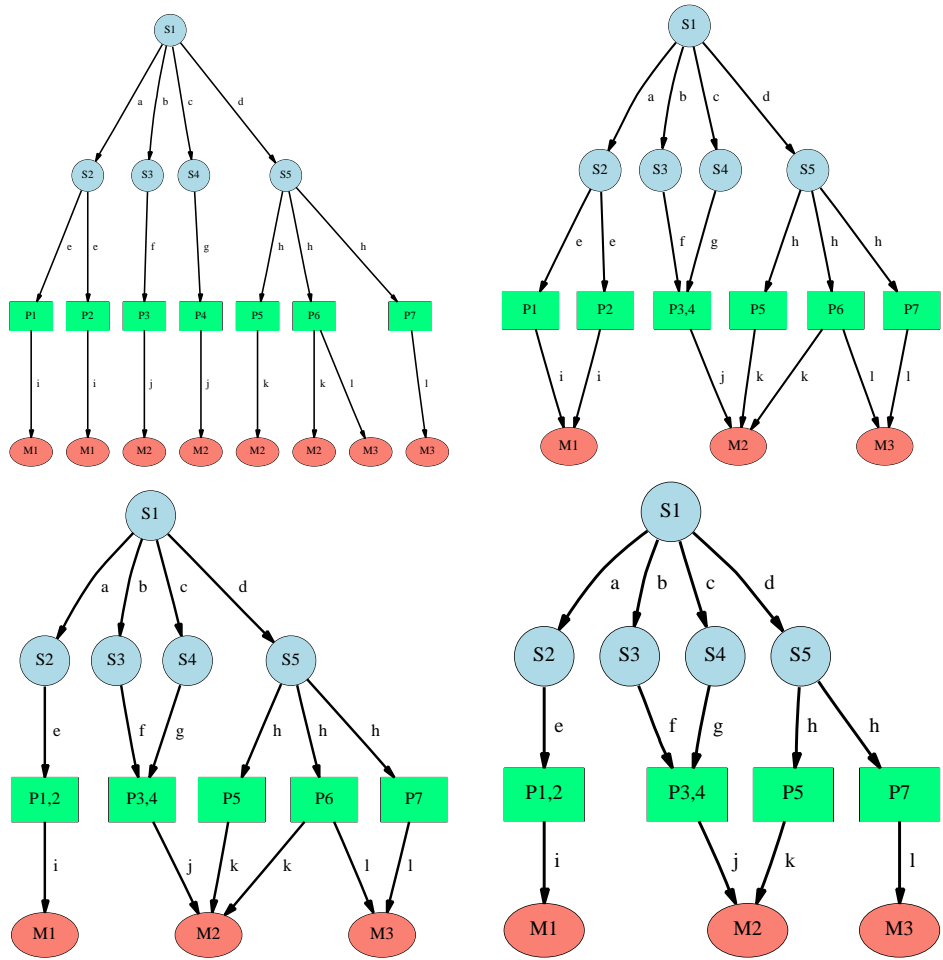

\centering
\begin{tabular}{cc}
\includegraphics[width=2.3in]{ex1}
\hspace{0.2in}
\includegraphics[width=2.3in]{ex2}
\end{tabular}
\begin{tabular}{cc}
\includegraphics[width=2.3in]{ex3}
\hspace{0.2in}
\includegraphics[width=2.3in]{ex4}
\end{tabular}
\caption{Four stages in extracting structure from a semistructured data source, by
the algorithm of~\cite{nestorov}. (top left) Original semistructured resource with
labeled and directed edges modeling interaction sequences. (top right)
Factorization of commonalities encountered in crawling. (bottom left)
A `minimal perfect typing' of the data. (bottom right) 
Final output of data mining algorithm, after modeling `multiple roles' ~\cite{nestorov}.}
\label{nestorov}
\end{figure}

\subsubsection*{Program Compaction}
The naive rendition of a PIPE model by the above mechanisms might result in
lengthy programs, with duplications of interaction sequences. Techniques for program
compaction are hence important. This topic has been studied extensively in the data mining
and semistructured modeling communities~\cite{dataontheweb,nestorov,tkde-structure}. 
Of particular relevance to PIPE is the algorithm 
of Nestorov et al.~\cite{nestorov} whose modeling of semistructure closely resembles 
our representation of an interaction sequence in terms of program variables. This algorithm
works by identifying graph constructs that could be factored, simplified, or approximated. 
Fig.~\ref{nestorov} describes four stages in a procedure for program compaction. The starting
point is the schema in Fig.~\ref{nestorov} (top left) obtained by a naive crawl of a site. 
Fig.~\ref{nestorov} (top right) factors commonalities encountered in crawling.
There are only three leaf nodes and the internal nodes {\tt P3} and {\tt P4} are collapsed because
they are really the same page.
Fig.~\ref{nestorov} (bottom left) is a `minimal perfect typing~\cite{nestorov}' of the data, which means that
the fewest internal nodes needed to describe the schema are used. In this example, {\tt P1} and
{\tt P2} are collapsed, not because they are the same but because they exhibit the same schema.
Both have an incoming edge labeled {\tt e} from the same type of page ({\tt S2}) and display an
outgoing edge labeled {\tt i} to the same type of page ({\tt M1}). While their contents
may not be the same, interaction sequences involving them can be compacted.
Care must be taken to ensure that any accompanying text with these nodes are not lost.
And finally, Fig.~\ref{nestorov} (bottom right) casts {\tt P6} as redundant for the purpose
of modeling
interaction sequences. The role of {\tt P6} in Fig.~\ref{nestorov} (bottom right) is to establish
connections from {\tt S5} to {\tt M2} and {\tt M3}, which are already embodied in
{\tt P5} and {\tt P7} respectively. Thus, we can remove {\tt P6}, once again after ensuring that
any contents of that node are suitably represented elsewhere. In~\cite{nestorov}, {\tt P6} is referred 
to as a node that exhibits `multiple roles.'
%and that the termination of the partial evaluator is not compromised by such approximations~\cite{jones-book}.

\subsubsection*{Miscellaneous Optimizations} 
Finally, the success of a personalization system relies on those finer touches that deliver
a compelling experience to the user. Options in this category are ad-hoc by nature and are not technically
modeling choices since they involve post-processing of the specialized
program. For instance, assume
that we personalize the automobile example in Fig.~\ref{pe2} with respect to the variables {\tt Honda} and {\tt 2001}.
This might produce a construct such as:
\begin{verbatim}
if (Green) {
        /* two empty code blocks */

        /* the first is empty because Honda and 2001 evaluated to true,
           but there were no green Honda cars made in 2001 */

        /* the second is empty because other models and other years were set
           to be evaluated to false */
}
\end{verbatim}
%This is due to the fact that in the interaction sequence {\tt make} and {\tt year} are modeled beneath
%{\tt color}. 
While semantically correct, such code blocks are useless for information presentation.
They can be perceived as dead-ends and safely omitted during web page reconstruction. It would also be
confusing to the user who clicks on `Green' and receives nothing (or an empty page) in return!

A second form of
optimization arises when partial evaluation results in a nested conditional with no {\tt else}
clauses:
\begin{verbatim}
if (Blue) {
   if (2001) {
     if (Honda) {
         /* something here */ 
     }
   }
}
/* nothing here */
\end{verbatim}
In such a case, we need to pay attention to how the simplified program is presented back to the user.
Forcing the user to continue clicking on items when there is only one choice at every level is undesirable.
Rather, we could just reveal to the user that according to his personalization criteria,
the only type of cars remaining are `Blue Honda 2001' and directly link to the items of information.
This example reinforces our idea that structural information is first-class information.
We are working on a customized partial evaluator that can perform such optimizations.

% 3. crawling clickable maps:
% 4. interacting with recommender systems: rec. sys. could be a f() within the model. Or, the aspect of recommendation could
%be "external" to PIPE, in that collaborative features correspond to program variables. "x" means.
% 5. browsing computed information: hooks,
% 6. modeling within a web page: virtual nodes, DTDs, naveen ashish, XTRACT,
% 7. program compaction: put 4 pictures

% generators of hierarchies (the one you found with Sammy)

\section{Application Case Studies}
\label{studies}

We now describe two applications that use PIPE to personalize collections
of web sites. They are presented in increasing order of complexity, as evidenced
by the forms of modeling they conduct (Table~\ref{whatismodeled}). In each
of these applications, we state the conceptual model of interaction sequences and
the specific choices made in modeling. Evaluation methodologies
are outlined after the descriptions. Since PIPE only specializes representations,
we are able to personalize even third-party sites by forming suitable representations.
More personalization systems designed with PIPE are described in~\cite{naren-ic,pipe-tochi}; we present only
two here for space considerations.

\begin{table}
\centering
\begin{tabular}{|lcl|} \hline
Congressional Officials &   & Modeling Site Structure \\
                        &   & Modeling within a Page\\
			& & \\
% Pigment Composition and Analysis & &  Modeling Site Structure\\
% 			&   & Information Integration \\
%                         &   & Browsing Computed Information\\
%			& & \\
Mathematical and Scientific Software & &  Modeling Site Structure\\
			&   & Interacting with Recommender Systems \\
 			&   & Information Integration \\
                         &   & Modeling within a Page\\
                         &   & Program Compaction\\
\hline
\end{tabular}
\caption{Modeling options used in the application case studies.}
\label{whatismodeled}
\end{table}

\subsection{Congressional Officials}
\label{politics}
Our first application customizes access to the
Project Vote Smart website (\url{http://www.vote-smart.org}), an independent resource
for information about United States governmental officials.
The site caters to people interested in
politicians' backgrounds, committee memberships, and positions
on major political issues. 
While Project Vote Smart reports on state and local governments as
well as the federal government, we focused only on the
congressional subsection of the site in our experiments.

The conceptual model of information-seeking involves browsing through the congressional
subsection to retrieve individual web pages of politicians. Interaction sequences at this
site consist of choices of state (e.g., California, Virginia, etc.), branch of 
congress (House or Senate), party (Democrat, Republican, or Independent), and 
district information (numbers of districts). The terminal information involved 540
home pages (for 100 Senate members and 440 House members) and resides at the 
ends of interaction sequences.

\begin{figure} 
\centering
\begin{tabular}{c}
\includegraphics[width=12.4cm,height=12.6cm]{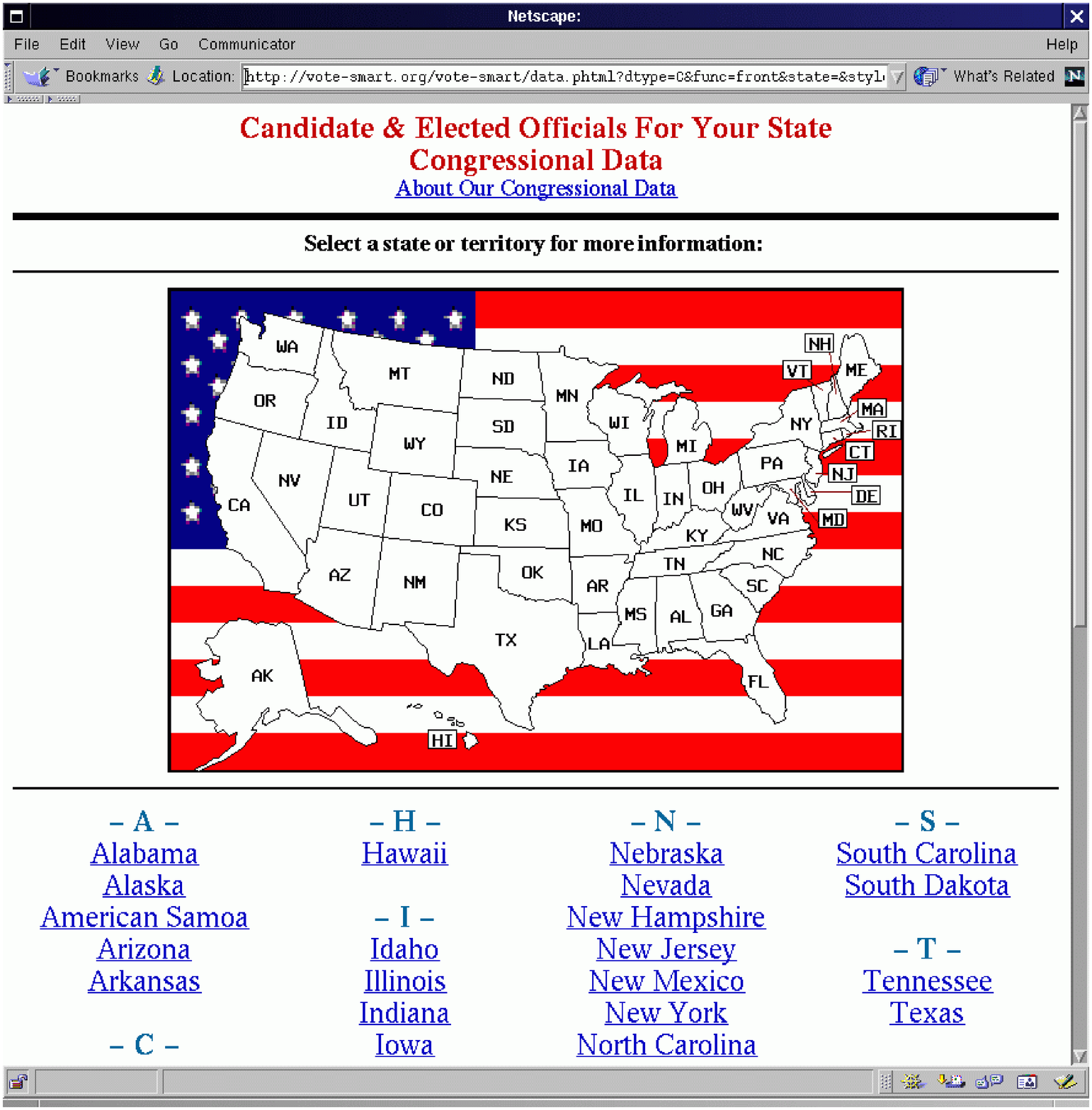}
\end{tabular} 
\begin{tabular}{cc}
\includegraphics[width=6.2cm,height=6.84cm]{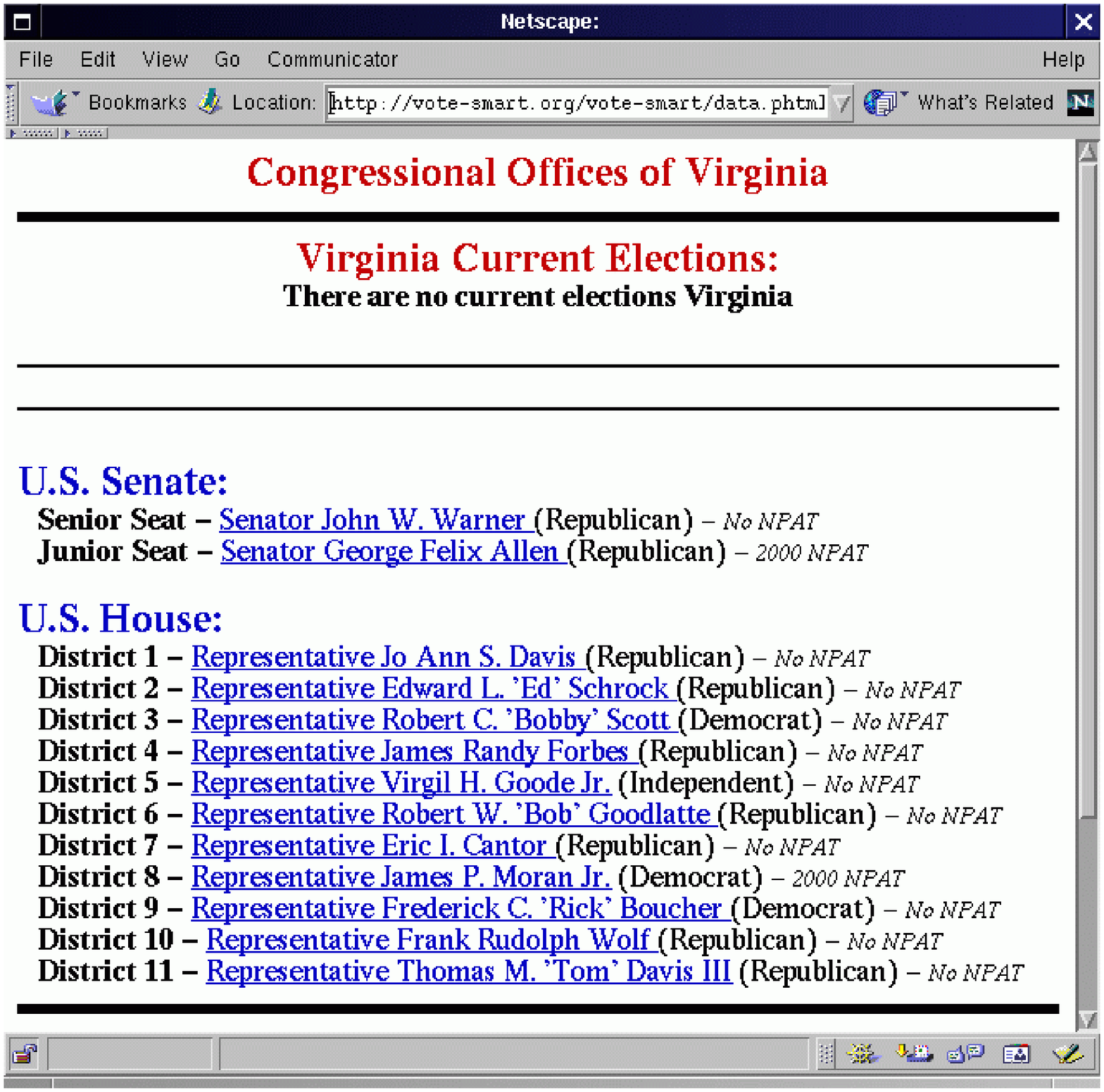}
\includegraphics[width=6.2cm,height=6.84cm]{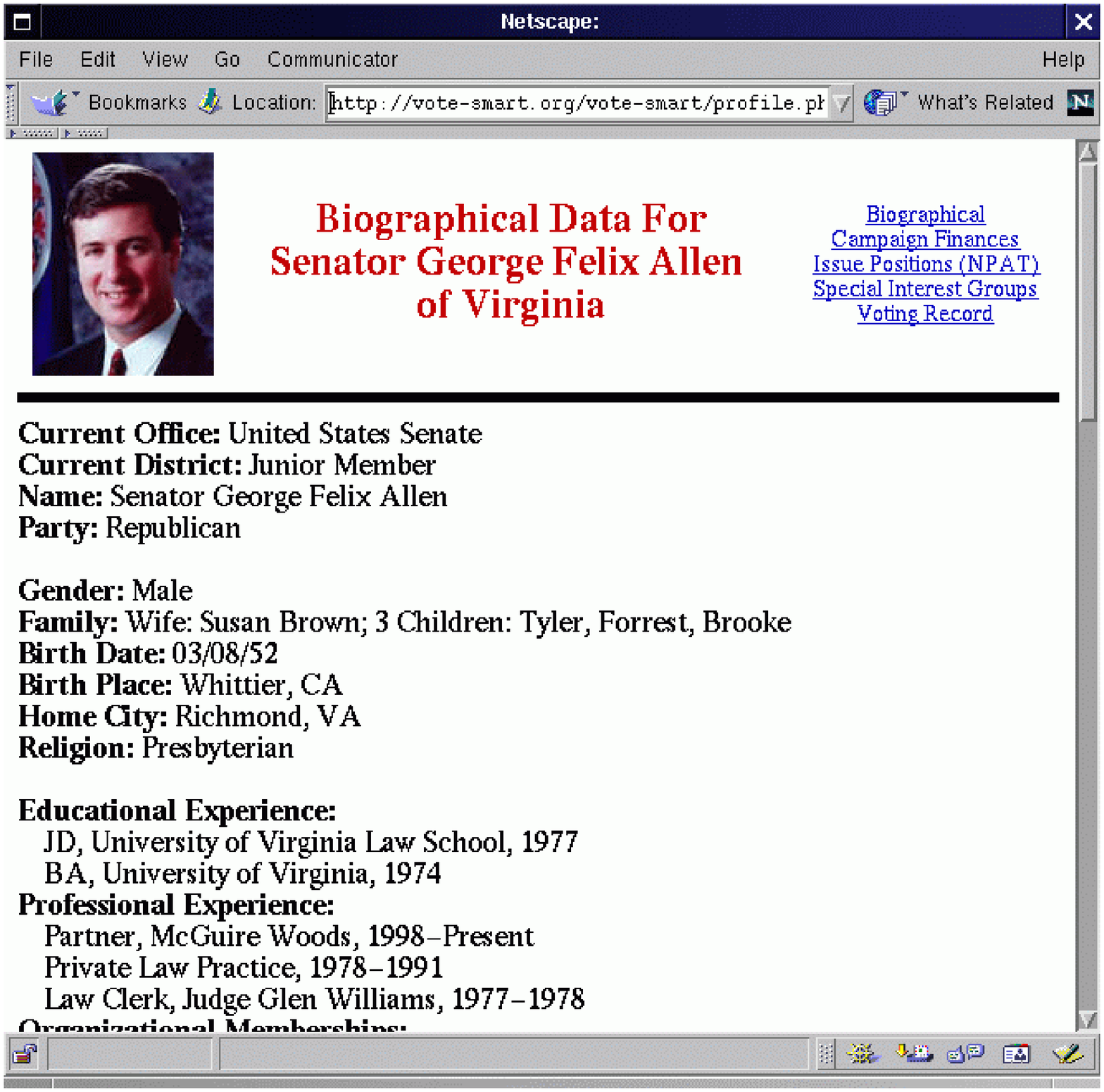}
\end{tabular} 
\caption{A typical interaction sequence at the Project Vote Smart web site. (top) Start page
for congressional officials. Making a selection of state at this level reaches a state-level
page (bottom left). Finally, individual politicians' web pages are accessed by making selections
at the state-level page (bottom right).}
\label{politicians1}
\end{figure}

Fig.~\ref{politicians1} describes a typical interaction sequence.
At the root congressional page (Fig.~\ref{politicians1} (top)),
users are directed to select a state of interest.
Selection of state transfers the
user to that particular state's web page
(Fig.~\ref{politicians1} (bottom left)).  
A state web page is semistructured, listing
both senators and representatives as well as their party, district
affiliations, and other associated information.  Finally, a user
arrives at a politician's webpage (Fig.~\ref{politicians1} (bottom right))
by making a selection at the state page.  Thus, the congressional
section of Project Vote Smart is three levels deep (with a two-step interaction
sequence).

Since many of the choices made
by the user in browsing through Project Vote Smart are independent
of each other (e.g., selecting Virginia as state does not imply a particular political party),
the site is highly amenable to personalization by partial evaluation. 
Currently the site hardwires interaction sequences in the order shown in Fig.~\ref{politicians1}.
We modeled the two-step interaction sequence (as shown in Fig.~\ref{politicians1}) as
actually a four-step interaction sequence by conducting a more detailed modeling of the
state-level page. In particular, the semistructure on state-level pages was abstracted
to yield independently addressable information about branch of congress, party, and district.

The site graph is not a balanced tree. For instance, every state has exactly two senators
but the number of representatives varies from 1 in South Dakota to 52 in California (this
is dependent on state population).  Our modeling of data at state pages 
expanded the original 3-level tree shown in Fig.~\ref{politicians1}  
consisting of 596 nodes (1 root page + 55 state pages + the 
previously mentioned 540 leaves of the tree) to 5 levels comprising
857 nodes (317 internal nodes + 540 leaf nodes).  This amounts to
a approximately 44\% percent explosion in the site schema.

%3111 lines of C code and 129 program variables that are addressable. 
The programmatic representation of the new site schema was in C and it
captured miscellaneous domain semantics about interaction at the site (e.g.,
if the user says `District 21,' he is referring to a Representative, not
a Senator). The partial evaluator C-Mix was used for this study.

\subsection{Mathematical and Scientific Software}
Our second application is a personalization system for recommending mathematical 
software on the web for scientists and engineers. Consider a scientist studying stress
in a helical spring; he formulates the problem mathematically in terms of
a partial differential equation (PDE) and proceeds to find software that can help
in solving his PDE. He uses a collection of three web sites to conduct his information-seeking
activity.

First, he accesses the GAMS (Guide to Available Mathematical Software)
cross-index of mathematical software ({\tt http://\hskip0ex gams.\hskip0ex nist.\hskip0ex gov}), a
tree-structured taxonomy that covers nearly 10,000 algorithms (from
over 100 software packages) for most areas of scientific software.
GAMS functions in an interactive fashion, guiding the user
from the top of a classification tree to specific modules as the
user describes his problem in increasing detail. During this process,
many important features of the software (e.g., `are you looking for a software to solve elliptic
problems?') are determined, from the user.
However at the ends of the interaction sequences at GAMS, there still exist several choices 
of algorithms for a specific problem.
Now, the scientist consults a recommender system or a performance database server
(for his category of scientific software) to pick an appropriate algorithm for his problem. 
An example is the PYTHIA recommender system for selecting solvers for PDEs~\cite{pythiaii}.
At this point, the scientist supplies additional information to the recommender such as
his performance constraints (on the time to solve his PDE). 
Systems like PYTHIA use previously archived performance data to arrive at recommendations such as
`Use the second-order 9-point finite differences code from the ELLPACK module.'
After such a recommendation, the scientist conducts the final step of downloading the 
recommended software module from repositories such as Netlib ({\tt http://\hskip0ex www.\hskip0ex netlib.\hskip0ex org}) 
housed at the Oak Ridge National Laboratory (ORNL) or other packages at the National Institute 
of Standards and Technology (NIST). The conceptual model involved the information flow from the GAMS 
site, to a repository such as Netlib, through a recommender such as PYTHIA. 

The choices made in GAMS will affect the choice of recommender which in turn affect the choice
of repository. This application thus presents an interesting information flow for modeling.
Since PIPE permits partial instantiation of the information flow, the scientist can
directly access a repository such as Netlib if he is sure of the specific software he needs. 

We modeled the entire GAMS web site, used the PYTHIA recommender (that addresses
software for the domain of PDEs), and established connections with individual software modules at the
various repositories. After an initial expansion of GAMS (e.g., by within-page modeling),
we applied the program compaction algorithm described in Section~\ref{options}.
Cross-references in GAMS and duplication of common module sets (which are now revealed
by our initial expansion) helped compress the site schema to 60\% of its original size.
In particular, the GAMS subtree relevant to describing PDEs provided for a 11\% compression.
There was no terminal information alongside intermediate nodes, and hence there was no
need for any special handling. PYTHIA's details are described in~\cite{pythiaii} and we conducted
a white-box modeling in PIPE to better associate program variables from GAMS with variables in
PYTHIA (one of the authors of this paper was also the co-designer of the PYTHIA recommender).
Finally, the step to reach individual software modules was a simple one-step interaction sequence 
leading to terminal information about the code (in FORTRAN) and its documentation. 
The entire composite program was represented in the CLIPS programming language~\cite{clips} and 
we employed its rule-based interface for partial evaluation. More modeling details on this case study
can be found in~\cite{pipe-gams}.

\subsection{Evaluation}
\label{eval}
We now describe procedures for evaluation. There are three possible types of evaluation:

\begin{enumerate}
\item Evaluating PIPE applications
\item Evaluating our modeling of information-seeking activities in PIPE
\item Evaluating PIPE
\end{enumerate}

The first type of evaluation is what is usually described in the literature
and there are many ways of conducting it. 
The accepted practice is to measure improvements in revenues,
site visits, and user satisfaction (e.g., via surveys). In~\cite{naren-ic},
we have described the evaluation of PIPE applications using traditional user 
interviews followed by statistical validation (they have yielded good results). Commercial
ventures such as {\it NetPerceptions} emphasize the scalability and speed-of-response of 
personalization systems. The second and third types of evaluation criteria highlight the role
of PIPE as a modeling methodology.
We concentrate on them since we have already described traditional user-response
evaluation of PIPE applications in~\cite{naren-ic}.
This section covers the evaluation
of modeling and Section~\ref{factor} helps identify shortcomings of the PIPE methodology
itself.

We evaluate a PIPE modeling by the extent to which
it allows users' information-seeking activities to be described as partial inputs.
This is in keeping with the view that PIPE's services are only as good as the
modeling conducted in it. If a faulty recommender system is modeled in PIPE, then no
amount of partial evaluation can provide satisfactory results.

Recall that our modeling was conducted with respect to a set of interaction sequences.
For evaluation purposes, we identified an independent `external examiner' 
model, which was also a set of interaction sequences. 
We then evaluated our PIPE modeling 
by the fraction of interaction sequences in the external examiner model that can 
be realized by an appropriate partial evaluation operation. We discounted optimizations such 
as described in Section~\ref{options} when determining the `unrealizable' 
interaction sequences.

In the first study, the
examiner model was obtained from users. They were provided
knowledge of the functional specification of our original conceptual
modeling, not its details. For instance, they
were told about the nature of structural and terminal information (and 
any functional dependencies among them), but not the exact interaction sequences that 
constitute the conceptual model. Formal methodologies for this activity are described
in~\cite{gannon-verify}.

We identified 25 user subjects who were predominantly graduate students from Virginia 
Tech (but not necessarily computer science majors). The ages of the subjects ranged from 
19 to 49, with the average age being 26. A majority of the subjects rated their 
computer and web familiarity as above average. All subjects acquainted themselves with the 
Project Vote Smart site by browsing for about ten minutes. Each subject was then asked to 
describe 1-2 personalization scenarios. Notice that these are different from `queries,'
as they specified constraints on interaction e.g., `I would like to browse by state,
and then I will make a choice of party, and then I would click any remaining hyperlinks
to browse the site.' 

In total, 32 interaction sequences were identified, of which 25 were 
realizable in our modeling. One of the unmodelable 
scenarios was `I would like to see all politicians who represent Los Angeles,' a request that 
was not faithful to our conceptual model. We do not discuss this further. The other six 
unmodelable scenarios are not shortcomings of our modeling, but rather shortcomings of the 
PIPE personalization methodology itself. They involved restructuring operations on interaction 
sequences that are not describable as partial evaluations. Section~\ref{factor} analyzes 
these in detail.

\begin {figure}
\centering
\begin {tabular}{|lp{3.5in}|}\hline
{\sc Problem \#28} & ${\left(w\,u_x\right)}_x + {\left(w\,u_y\right)}_y = 1,$\\
 & where $w = \left\{ \begin {array}{l}
                               \alpha,\,\,\, {\rm if}\,\,\, 0 \leq x,y \leq 1\\
                               1,\,\,\, {\rm otherwise.}
                              \end {array}
                      \right.$\\
{\sc Domain} & $\left[-1,1\right] \times \left[-1,1\right]$\\
{\sc BC} & $u = 0$\\
{\sc True} & unknown\\
{\sc Operator} & Self--adjoint, discontinuous coefficients\\
{\sc Right Side} & Constant\\
{\sc Boundary Conditions} & Dirichlet, homogeneous\\
{\sc Error Constraint} & 1.0E-05 \\
{\sc Time Constraint} & 60s
\\\hline
\end {tabular}
\caption {A problem from the examiner model for the second case study.}
\label{pde-scenario}
\end {figure}

For the second study, the examiner model was derived from a benchmark set of problems that are used
in mathematical software evaluation (the set is described in~\cite{pythiaii}). Each of these problems
describes scenarios in terms of features of the PDE problem (e.g., is it Laplace?, is it Helmholtz?)
any constraints on its solution (e.g., relative error should be $< 10^{-9}$), and any restrictions
on software modules (e.g., `I would like to use the package NAG' or `ELLPACK modules are prefered.'). 
Fig.~\ref{pde-scenario} describes an example scenario that places constraints on the type of software
to be used (for instance, it should be applicable to `Dirichlet' problems) and the basis for
recommendation (namely, that it should satisfy the time and error constraints specified). This scenario does
not give any preferences for software modules or packages. Such mathematical descriptions are translated
into parameters for personalization (a process is described in~\cite{pythiaii}).
The examiner model comprised of 35 such interaction sequencs, of which all are modelable. More details on this
case study can be obtained from~\cite{pipe-gams}.
\section{Discussion}
\label{discuss}
\subsection{Related Research}

As a systematic methodology for personalization, PIPE is a unique research project.
Most research on personalization emphasizes the nature of information being 
modeled~\cite{specissue,phoaks} (content-based~\cite{fab} versus 
collaborative~\cite{adomavicius01expert-driven,
ira,grouplens,Siteseer}), the level at 
which the 
personalized information is targeted (is it by user~\cite{manber}, by topic~\cite{adaptive-sites} 
or for everybody~\cite{holland-hill,footprints}), or the specific algorithms that are involved in
making recommendations. 
%Evaluation metrics have also been influenced by these viewpoints.

In contrast, PIPE models interaction with an information system as the basis for personalization.
Most of recommender systems research can be viewed as modeling options for PIPE. The systems that
make distinctions among targeting constitute making different assumptions on the possible set of
interaction sequences. They can hence be tied to requirements analysis, as described 
in~\cite{pipe-tochi}. Systems that conduct web usage mining~\cite{cacm-jaideep,cacm-mulvenna} also
address the earlier parts (and sometimes, later parts~\cite{cacm-myra})
of the personalization system design lifecycle, and can be viewed as methodologies 
to suggest and refine interaction sequences.

Other connections to information systems research can be made by observing that PIPE
contributes both a way to model information-seeking activities as well as a closed transformation
operator for personalization i.e.,
partial evaluation. RABBIT~\cite{rabbit} is an early interactive 
information retrieval methodology that resembles PIPE in this respect. It proposes
the model of `retrieval by reformulation' to address the mismatch between 
how an information space is organized and how a particular user forages in it. Several closed transformation
operators are provided in RABBIT to enable the user to specify and realize information-seeking goals.
Like RABBIT, PIPE
assumes that `the user knows more about the generic structure of the [information space] than [PIPE]
does, although [PIPE] knows more about the particulars ([terminal information])~\cite{rabbit}.' For
instance, personalization by partial evaluation is only as effective as the ease with
which program variables could be set (on or off) based on information supplied by the user.
Unlike RABBIT, PIPE 
emphasizes the modeling of an information
space as well as an information-seeking activity in a unified programmatic representation.
Its single transformation operator is expressive enough to simplify a variety of
interaction sequences. 

%In addition, PIPE achieves mixed-initiative interaction without providing it as a specific
%function or requiring any explicit effort on the part of the user.

%The Scatter-Gather~\cite{scatter-gather} and Dynamic Taxonomies~\cite{tkde-navigation} projects 
%resemble PIPE in that they contribute closed transformation operators for retrieval and navigation,
%respectively. However, they do not emphasize representation as much as PIPE (or RABBIT)
%and adopt traditional modeling methodologies. Scatter-Gather introduces two operations 
%(Scatter and Gather) that are used for clustering and declustering documents. 
%The Dynamic Taxonomies project assumes a conceptual model of `selective thinning of an infobase,'
%and provides several set-oriented operations for navigation. 

The closed nature of transformation operators is central to interactive modes of information seeking,
as shown in projects such as Scatter-Gather~\cite{scatter-gather} and Dynamic Taxonomies~\cite{tkde-navigation}.
PIPE is novel in that it contributes a transformation operator for {\it representations of interactions} in
information spaces, and does not transform documents or web pages directly.

The `larger' approach to personalization taken in this paper is reminiscent of the integration
of task models in software design~\cite{intTaskObj}. Typically such integration has utilized 
object oriented methodologies and symbolic modeling approaches e.g., UML. This idea has been used for designing
personalization systems as well~\cite{li-catalog,schwabe2,human1,schwabe1}. However, in all of these projects,
personalization is introduced a function from the conceptual design stage. PIPE's
support for personalization, on the other hand, is built into the programmatic model of the information 
space and doesn't require any special handling.

\subsection{When PIPE does not Work: Reasoning about Representations}
\label{factor}

We now address limitations and some fundamental implications of the PIPE
methodology. We will explain why the six unmodelable interaction sequences in Section~\ref{politics}
are shortcomings of the PIPE methodology itself.
Let us first recall why examples such as Fig.~\ref{pe2} and the other application
study in Section~\ref{studies} work so well: Information-seeking activities in
these scenarios were describable as partial inputs in the modeling. Since the modeling
was parameterized in terms of program variables, another way to explain the
success of these applications is to say that `the representation of the
information space is factored in terms of structural information.' 

This suggests that it will be useful to understand how information spaces are factored,
in general. If the representation of the information space is not factored at all, it means that no program 
variables are available to be turned on or off and hence the space is not 
personalizable by PIPE. What is counterintuitive is that `too much factoring' could 
also render PIPE inapplicable or useless.

Consider our automobile example from Fig.~\ref{pe2} in Section~\ref{example}. 
It is reproduced in Fig.~\ref{newfactor} (right) with the addition of
some line numbers (to denote particular points in the program).
We can think of this as a factorization in terms of variables such as 
{\tt Blue} and {\tt Honda}, which in turn allow us to describe user requests.
The left part of Fig.~\ref{newfactor} describes an alternative
factorization of the same information space. In this case, the program variables and their
connections are stored in a `structure table' and an explicit generator is used to construct
the information space in Fig.~\ref{newfactor} (right). For instance, the structure table associates
the {\tt Blue} program variable as the condition that gets us from line 1 to line 2 in the
modeling. We can think of the structure table as modeling the site graph and the generator
as a depth-first search (DFS) algorithm that walks the site graph to construct the information space.

\begin{figure}
\centering
\begin{tabular}{lll}
\begin{tabular}{|l|l|l|} \hline
\multicolumn{3}{|c|}{\bf Structure Table} \\ \hline
From & To & Program \\ 
Line & Line & Variable \\ \hline
1 & 2 & Blue \\ 
1 & 3 & Red \\
2 & 4 & 2001 \\
2 & 5 & 2000 \\
$\cdots$ & $\cdots$ & $\cdots$ \\
\hline
\end{tabular} & \large{$\times$ Site Generator (e.g., DFS)} = & 
\begin{tabular}{|l|} \hline
{\tt L1:}\\
{\tt if (Blue)} \\
{\tt L2:}\\
\,\,\,\,{\tt if (2001)} \\
{\tt L4:}\\
\,\,\,\,\,\,\,\,{\tt if (Honda)} \\
\,\,\,\,\,\,\,\,\,\,\,\,{$\cdots \cdots \cdots$} \\
\,\,\,\,\,\,\,\,{\tt else if (Toyota)} \\
\,\,\,\,\,\,\,\,\,\,\,\,{$\cdots \cdots \cdots$} \\
\,\,\,\,{\tt else if (2000)} \\
{\tt L5:}\\
\,\,\,\,\,\,\,\,{$\cdots \cdots \cdots$} \\
{\tt else if (Red)} \\ 
{\tt L3:}\\
\,\,\,\,{\tt if (2001)} \\
\,\,\,\,\,\,\,\,{$\cdots \cdots \cdots$} \\
\,\,\,\,{\tt else if (2000)} \\
\,\,\,\,\,\,\,\,{$\cdots \cdots \cdots$} \\
\hline
\end{tabular}
\\
\end{tabular}
\caption{An example of a over-factored information space for personalization by partial evaluation. (left)
Modeling the generation of an information space. (right) Modeling the interaction in an information
space.}
\label{newfactor}
\end{figure}

Rather than think of the left part of Fig.~\ref{newfactor} as the {\it generator of an information space} and contrast
it with the right side (which describes it directly), let us temporarily think of both the left and 
right sides of Fig.~\ref{newfactor} as {\it alternative representations} of the same 
information space. The word `representation' does not
imply the mechanical aspect of constructing the information space (left of Fig.~\ref{newfactor})
or the interaction with the information space (right of Fig.~\ref{newfactor}). Since partial evaluation
merely specializes programs, it doesn't pay any attention to whether the program is meant to represent
interaction or generation. By losing this distinction (temporarily), we will be able to reason
about representations in general.

In Fig.~\ref{pe2}, we personalized the representation w.r.t. `2001'; the result
was shown in Fig.~\ref{pe2} (right).
Let us reconsider how we will address this request with the new design
shown in Fig.~\ref{newfactor} (left). 
We cannot specify this input to the DFS algorithm since it is not
parameterized in terms of specific variables like {\tt 2001.} The DFS is meant to work for all types of
trees and graphs, not just an automobile browsing hierarchy. We also cannot specify {\tt 2001} in terms of
the structure table since we have to manually readjust the line numbers to conform to the request.
The only way we can obtain the same result as in
Fig.~\ref{pe2} is to change the structure table in Fig.~\ref{newfactor} completely to reflect the
tree shown in Fig.~\ref{pipe-illustrate} (right). But by then, we have done most of the work needed
for personalization! In fact, the personalization request is no longer describable as partial evaluation,
but as a {\it complete evaluation} (specifying all arguments).
We say that such a design is {\it over-factored}, for the given
information-seeking activity.

Attempting to use an over-factored representation (for the type of information-seeking
activities in Fig.~\ref{pe2}) appears fruitless.
The reason is that over-factorization divorces two crucial elements
out, which really have to interplay for partial evaluation to be beneficial. Fig.~\ref{newfactor} (left) is like
two sides of the PIPE coin separated: the structure table contains the structural information (with which
we connect user requests) and the DFS contains the logic flow (which is simplified by partial
evaluation for the user). Neither is useful in PIPE without the other and yet they cannot be represented
distinctly. {\it This is why over-factorization is not desirable.}

It is important to note that an information system design is not just over-factored, it is over-factored
for a particular information-seeking activity. For instance, we can give an example of an information-seeking
activity for which the design in Fig.~\ref{newfactor} (left) is factored `just right.' Consider the following 
user who walks into the automobile dealership: 

\begin{descit}
\noindent
{\bf Buyer:} I am here to buy a car. Ask me the questions for year, model, and color, in that order.
\end{descit}
\
In this case, the user does not want a personalized information space for browsing. Rather, he is seeking
to personalize the {\it generation} of an information space. Our original modeling in Fig.~\ref{newfactor} (right) cannot
handle this situation. It can let the user give values out of turn, but it can't change the default order in which the
questions are asked. We say that the design in Fig.~\ref{newfactor} (right) is {\it under-factored} (for this activity).
However, the design in
Fig.~\ref{newfactor} (left) can accommodate it, if the site generator
can take arguments such as what the first level of the hierarchy should be, what the
second level should be, and so on. Presumably such a generator would walk the tree described by the structure table
and restructure it based on the arguments. In this case, we can still use partial evaluation for requests such as:

\begin{descit}
\noindent
{\bf Buyer:} I am here to buy a car. I don't care in what order you ask the questions, but the second
question should be about year.
\end{descit}
\noindent
(It is a different issue if such scenarios are likely. For now, we are only exploring the PIPE concept theoretically.)
After this information space is generated, we still have the option of  re-representing the generated information space 
in our usual manner and conducting personalization by partial evaluation. 
We can thus state the following three definitions:
\begin{descit}{}
\noindent
A representation $\mathcal{I}$ of an information space is well-factored for an information-seeking 
activity $\mathcal{G}$ if all interaction sequences in $\mathcal{G}$ can be realized by 
partial evaluations of $\mathcal{I}$. In this case, we also say that $\mathcal{I}$ is personable 
for $\mathcal{G}$.

A representation $\mathcal{I}$ of an information space is over-factored for an information-seeking 
activity $\mathcal{G}$ if all interaction sequences in $\mathcal{G}$ can be realized by
complete evaluations of $\mathcal{I}$. In this case, we also say that $\mathcal{I}$ is not personable 
for $\mathcal{G}$.

A representation $\mathcal{I}$ of an information space is under-factored for an information-seeking 
activity $\mathcal{G}$ if no interaction sequences in $\mathcal{G}$ can be realized by partial (or complete)
evaluations of $\mathcal{I}$. In this case, we also say that $\mathcal{I}$ is not 
personable for $\mathcal{G}$.
\end{descit}

%\begin{figure}
%\centering
%\begin{tabular}{cc} 
%\includegraphics[width=3in]{venndiagram}
%\end{tabular}
%\caption{Overlap between interactions sequences that are describable by complete evaluation and those
%that are describable by partial evaluation.}
%\label{venn}
%\end{figure}

Thus, a given representation could be well-factored for one information-seeking activity but over-factored
for another. Fig.~\ref{newfactor} (left) is well-factored for generation but over-factored for interaction.
Fig.~\ref{newfactor} (right) is well-factored for interaction but over-factored for 
users who employ the color-year-model motif diligently (and completely). 

The 6 unmodelable scenarios in Section~\ref{politics} involved requests such as `I would like to have the choice
of party as the first level of the hierarchy, the choice of state as the second level.' Our design was obviously
under-factored for such interaction sequences. We can define the {\it personability} of a representation as the
fraction of interaction sequences in a (external examiner) model that are describable as partial evaluations. For the
external examiner model described in Section~\ref{eval}, the personability of the PIPE modeling 
(presented in Section~\ref{politics}) is thus 25/32.

Notice that all of these statements assume that the model for transforming representations is {\it partial
evaluation.} There are other program-transformation techniques which might be able to address the
unmodelable requests above, but PIPE only provides partial evaluation as the operator for personalization.
Our statements should only be interpreted in the context of personalization by partial evaluation.

In practice, the decision of choosing a factoring will depend on which situations are more
likely and also the composition of the space of interaction sequences $\mathcal{G}$. It is acceptable to have
some interaction sequences that involve complete evaluation, as long as they are 
a small fraction of the total number of interaction sequences. 

Thus far, we have fixed the representation and analyzed the information-seeking activities for which it
was over-factored, the ones for which it was under-factored, and so on. This is the designer's viewpoint.
For a given site design, it allows the designer to pose questions such as `What are the information-seeking activities 
for which my site is personable?'

An alternate viewpoint is user-driven. Given an information-seeking activity, the user asks `What sites are
most personable for my activity?' This allows the user to take different site designs (along with representations), 
analyze them w.r.t. a conceptual model of information seeking, and rank them in order of personability. For instance,
consider again the external examiner model described in Section~\ref{eval} for the politicians case study. One information
system design was described in Section~\ref{politics}. The personability of this design is, as stated earlier, 25/32.
Seven interaction sequences were not modelable.
Another information system design is the representation in Fig.~\ref{newfactor} (left). The personability of this
design is 6/32. While it accommodates six of the seven sequences, it is no longer personable for the original
25 sequences! This is because those 25 sequences are now describable as complete evaluations, which also violate
the partial evaluation model! Thus, both over-factorization and under-factorization lead to unpersonable 
information spaces. We hypothesize that the most interesting representations are in between. 

An open research issue is if we have to cross the barrier 
from interaction to generation to arrive at over-factored representations. 
\section{Concluding Remarks}
\label{conclu}
This paper makes several major contributions. We have presented a novel modeling 
methodology for information personalization. PIPE enables the view of personalization
as specializing representations. It models interactions with information systems 
and uses partial evaluation to simplify the interactions. PIPE also contributes a novel
evaluation criterion for information system designs. It relates personalization to
the way an information system design is factored. This has implications for how web 
applications are developed and deployed~\cite{autoweb}. Many web sites today are based on
the generator model; the results in this paper indicate that they might not be 
directly personable for interaction scenarios (under partial evaluation). 

Our modeling makes very weak assumptions on the nature of interactions with
information systems. While we have covered only web sites (and collections of web sites)
in this paper, any information system technology that affords the notion of interaction
sequence or the idea of factorization
can be studied on similar lines. This especially applies to designs for voice-activated
systems (e.g., VoiceXML), directory access protocols (e.g., LDAP), information systems
that provide a dialog model of interaction, and models for organizing digital libraries (e.g.,
5S).

We plan to extend the PIPE methodology in several directions. We would like to extend the
modeling methodology to address earlier aspects of the personalization system design life cycle, such 
as requirements gathering, verification, and validation. First steps toward this goal are
described in a companion paper~\cite{pipe-tochi}. Another important direction of
future work involves modeling {\it context} in personalization systems. The programmatic modeling
provided in PIPE suggests that context can be usefully viewed as partial information. We believe
that more sophisticated forms of modeling partial information will be needed for describing context, besides
values for program variables.
We are also interested in relaxing our
assumptions of bounded sequences that have separable structural and terminal parts. This will
allow us to address other information-seeking activities such as social network navigation.
In addition, we are investigating program transformation techniques 
that can help reason about terminal information
(e.g., program slicing~\cite{slicing}),
in addition to structural information.

Our long-term goal is to develop a theory of reasoning about representations of information
spaces. This will allow us to formally study the design and implementation of information systems
in terms of the representations they employ.

\section*{Acknowledgements}
Many of the ideas in this paper were developed during the
Spring 2001 offering of the CS 6604 course (on recommender systems and personalization) at
Virginia Tech. We acknowledge helpful discussions with 
Jack Carroll, Marcos Gon\c{c}alves, Dennis Kafura, Priya Lakshminarayanan, Dick Nance,
Manuel Perez, and Mary Beth Rosson. 
Rob Capra helped establish connections between PIPE and mixed-initiative interaction and provided
ideas for evaluating the modeling of personalization systems.
Ed Fox suggested the usage of `structural' and `terminal' information to qualify interaction
sequences. Comments from several anonymous referees helped improve the presentation of the article.

\bibliographystyle{plain}
\bibliography{paper}

\end{document}